\PassOptionsToPackage{unicode}{hyperref}
\PassOptionsToPackage{hyphens}{url}
\PassOptionsToPackage{dvipsnames,svgnames,x11names}{xcolor}
\documentclass[
  12pt]{article}

\usepackage{amsmath,amssymb}
\usepackage{iftex}
\ifPDFTeX
  \usepackage[T1]{fontenc}
  \usepackage[utf8]{inputenc}
  \usepackage{textcomp} % provide euro and other symbols
\else % if luatex or xetex
  \usepackage{unicode-math}
  \defaultfontfeatures{Scale=MatchLowercase}
  \defaultfontfeatures[\rmfamily]{Ligatures=TeX,Scale=1}
\fi
\usepackage{lmodern}
\ifPDFTeX\else  
\fi
\makeatletter
\@ifundefined{KOMAClassName}{% if non-KOMA class
  \IfFileExists{parskip.sty}{%
    \usepackage{parskip}
  }{% else
    \setlength{\parindent}{0pt}
    \setlength{\parskip}{6pt plus 2pt minus 1pt}}
}{% if KOMA class
  \KOMAoptions{parskip=half}}
\makeatother
\usepackage{xcolor}
\makeatletter
\ifx\paragraph\undefined\else
  \let\oldparagraph\paragraph
  \renewcommand{\paragraph}{
    \@ifstar
      \xxxParagraphStar
      \xxxParagraphNoStar
  }
  \newcommand{\xxxParagraphStar}[1]{\oldparagraph*{#1}\mbox{}}
  \newcommand{\xxxParagraphNoStar}[1]{\oldparagraph{#1}\mbox{}}
\fi
\ifx\subparagraph\undefined\else
  \let\oldsubparagraph\subparagraph
  \renewcommand{\subparagraph}{
    \@ifstar
      \xxxSubParagraphStar
      \xxxSubParagraphNoStar
  }
  \newcommand{\xxxSubParagraphStar}[1]{\oldsubparagraph*{#1}\mbox{}}
  \newcommand{\xxxSubParagraphNoStar}[1]{\oldsubparagraph{#1}\mbox{}}
\fi
\makeatother

\usepackage{longtable,booktabs,array}
\usepackage{calc} % for calculating minipage widths
\usepackage{etoolbox}
\makeatletter
\patchcmd\longtable{\par}{\if@noskipsec\mbox{}\fi\par}{}{}
\makeatother
\IfFileExists{footnotehyper.sty}{\usepackage{footnotehyper}}{\usepackage{footnote}}
\makesavenoteenv{longtable}
\usepackage{graphicx}
\makeatletter
\def\maxwidth{\ifdim\Gin@nat@width>\linewidth\linewidth\else\Gin@nat@width\fi}
\def\maxheight{\ifdim\Gin@nat@height>\textheight\textheight\else\Gin@nat@height\fi}
\makeatother
\setkeys{Gin}{width=\maxwidth,height=\maxheight,keepaspectratio}
\makeatletter
\def\fps@figure{htbp}
\makeatother

\makeatletter
\@ifpackageloaded{caption}{}{\usepackage{caption}}
\AtBeginDocument{%
\ifdefined\contentsname
  \renewcommand*\contentsname{Table of contents}
\else
  \newcommand\contentsname{Table of contents}
\fi
\ifdefined\listfigurename
  \renewcommand*\listfigurename{List of Figures}
\else
  \newcommand\listfigurename{List of Figures}
\fi
\ifdefined\listtablename
  \renewcommand*\listtablename{List of Tables}
\else
  \newcommand\listtablename{List of Tables}
\fi
\ifdefined\figurename
  \renewcommand*\figurename{Figure}
\else
  \newcommand\figurename{Figure}
\fi
\ifdefined\tablename
  \renewcommand*\tablename{Table}
\else
  \newcommand\tablename{Table}
\fi
}
\@ifpackageloaded{float}{}{\usepackage{float}}
\floatstyle{ruled}
\@ifundefined{c@chapter}{\newfloat{codelisting}{h}{lop}}{\newfloat{codelisting}{h}{lop}[chapter]}
\floatname{codelisting}{Listing}

\makeatother
\makeatletter
\@ifpackageloaded{caption}{}{\usepackage{caption}}
\@ifpackageloaded{subcaption}{}{\usepackage{subcaption}}
\makeatother

\ifLuaTeX
  \usepackage{selnolig}  % disable illegal ligatures
\fi
\usepackage[]{natbib}
\usepackage{bookmark}

\IfFileExists{xurl.sty}{\usepackage{xurl}}{} % add URL line breaks if available
\hypersetup{
  pdftitle={Title},
  pdfauthor={Author 1; Author 2},
  pdfkeywords={3 to 6 keywords, that do not appear in the title},
  colorlinks=true,
  linkcolor={blue},
  filecolor={Maroon},
  citecolor={Blue},
  urlcolor={Blue},
  pdfcreator={LaTeX via pandoc}}

\newcommand{\anon}{1}

\allowdisplaybreaks[4]

\usepackage{comment}
\usepackage{setspace}
\usepackage{algorithm}
\usepackage{algpseudocode}

\newcommand{\indep}{\mathop{\perp\!\!\!\perp}}

\newcommand{\bld}{\boldsymbol}

\newtheorem{thm}{Theorem}

\newtheorem{prp}{Proposition}

\newtheorem{assu}{Assumption}
\newtheorem{rem}{Remark}

\begin{document}

\def\spacingset#1{\renewcommand{\baselinestretch}%
{#1}\small\normalsize} \spacingset{1}

%%%%%%%%%%%%%%%%%%%%%%%%%%%%%%%%%%%%%%%%%%%%%%%%%%%%%%%%%%%%%%%%%%%%%%%%%%%%%%

\if1\anon
{
  \title{\bf Estimating the average treatment effect under limited overlap via Polynomial Approximation and Extrapolation}
  \author{Shunichiro Orihara\thanks{Email:\ orihara@tokyo-med.ac.jp}\ \ \ and\ \ Sho Komukai\\
    Department of Health Data Science, Tokyo Medical University\\
    Fan Li\\
    Department of Biostatistics, Yale School of Public Health}
  \maketitle
} \fi

\if0\anon
{
  \bigskip
  \bigskip
  \bigskip
  \begin{center}
    {\LARGE\bf Title}
\end{center}
  \medskip
} \fi

\bigskip
\begin{abstract}
Estimating the average treatment effect (ATE) remains a fundamental challenge in observational studies in the presence of poor or limited covariate overlap.
Although the inverse probability weighting (IPW) estimator is a widely used approach for estimating the ATE, its performance can deteriorate substantially when overlap is limited, often resulting in increased finite sample bias and unreliable confidence intervals.
One common strategy is to shift attention from the original target estimand, the ATE, to alternative estimands such as a class of weighted ATEs that are less sensitive to extreme propensity scores; however, doing so changes the scientific question of interest.
In this manuscript, we propose a novel ATE estimator that preserves the original target estimand, the ATE,  while improving robustness to limited overlap.
A key idea is that this class of estimands can be represented by a polynomial function of a hyperparameter characterizing the estimands.
Exploiting this structure, the proposed method computes IPW estimators for a sequence of such estimands, models these estimates using a polynomial regression, and extrapolates to recover the ATE.
We show that the estimator has consistency and asymptotic normality under weaker overlap conditions than required for the standard IPW estimator.
Simulation studies demonstrate that the proposed method improves estimation accuracy and interval performance in settings with limited overlap.
In addition to its theoretical and empirical advantages, the proposed approach has a clear interpretation and is easy to implement using standard statistical software.
\end{abstract}

\noindent%
\vfill

\newpage
\spacingset{1.8} % DON'T change the spacing!

\section{Introduction}

The propensity score plays a central role in causal inference with observational data by providing a summary of measured pre-treatment covariates that can be used to adjust for systematic differences between treatment groups \citep{rosenbaum1983central}. A canonical target of inference is the average treatment effect (ATE), defined as the contrast between the mean potential outcomes had all members of the target population received treatment versus control \citep{imbens2015causal}. Among propensity score methods, inverse probability weighting (IPW) reweights the observed data to construct a pseudo-population in which the distributions of measured covariates are balanced between treatment groups. Augmented inverse probability weighting (AIPW) additionally incorporates an outcome regression and, under suitable regularity conditions, can improve efficiency while remaining consistent if either the propensity score model or the outcome model is correctly specified.

Although the IPW estimator is straightforward to implement and enjoys desirable large sample properties, including consistency and asymptotic normality under standard regularity conditions \citep{zhou2020psweight,chesnaye2022introduction}, its performance can deteriorate sharply under \emph{poor} or \emph{limited overlap}. In particular, propensity scores close to $0$ or $1$ generate large inverse probability weights, allowing a small number of observations to dominate the estimator and resulting in high variance and unstable inference \citep{kang2007demystifying,li2019addressing,heiler2021valid}. One widely used design-based strategy is to redefine the target population and, consequently, the causal estimand. A prominent example is the average treatment effect for the overlap population (ATO) \citep{li2018balancing}, which emphasizes individuals with propensity scores near $0.5$ and continuously downweights those toward the boundaries. The ATO can therefore be interpreted as the causal effect among individuals in clinical equipoise \citep{rizk2025and}. The beta weight family \citep{matsouaka2024causal} generalizes this idea to a continuum of weighted average treatment effects. On the other hand, propensity score trimming adopts a related principle through discrete rather than continuous exclusion. Specifically, \citet{crump2009dealing} restrict the analysis to individuals whose estimated propensity scores lie within a prespecified interval and characterize overlap subpopulations for which treatment effects can be estimated more precisely. Although hard trimming is intuitive and easy to implement \citep{sturmer2010treatment,zhou2020psweight}, its nonsmooth, data-dependent inclusion rule complicates inference for the resulting target population. \citet{yang2018asymptotic} address this issue by replacing the hard trimming indicator with a smooth approximation, thereby obtaining asymptotically linear estimators and enabling resampling-based inference. More recently, \citet{khan2025doubly} broaden the trimming criterion beyond the propensity score by accounting jointly for treatment assignment probabilities and conditional outcome variances. They develop doubly robust, heteroscedasticity-aware procedures that permit valid inference for the resulting trimmed subpopulations under flexible nuisance function estimation. Relatedly, \citet{barnard2026framework} formalize the choice among alternative estimands by considering estimator variance, statistical bias, and estimand mismatch relative to the ATE.

These modified estimands are compelling when their induced target populations align with the scientific question. Their improved stability, however, is obtained by redefining the research question. In certain applications, the target population is determined a priori by scientific or policy considerations, and interest lies in the effect of treating all eligible individuals rather than only those in regions of strong overlap. In such settings, the ATE can still remain the substantively relevant estimand, and limited overlap represents an inherent, inferential challenge rather than, by itself, a reason to replace the target. Under this broad theme, several methods instead use trimming as a regularization device while continuing to target the original ATE. \citet{ma2020robust} develop bias-corrected inference for trimmed IPW estimators and select the trimming threshold by minimizing an empirical approximation to asymptotic mean squared error. \citet{chaudhuri2025heavy} instead trim extreme IPW contributions, rather than only extreme propensity scores, and correct the resulting trimming bias to obtain robust inference for the ATE under heavy tails. Taking a sensitivity analysis perspective, \citet{ma2025sensitivity} construct worst-case bounds for trimming bias under explicit assumptions governing extrapolation from regions of adequate overlap. These developments underscore the central difficulty. That is, attenuating observations in poorly supported regions improves stability, but recovering the original ATE requires information about precisely those regions.

Motivated by this tension, we propose a different use of alternative weighted estimands. Rather than selecting a single WATE as a replacement for the ATE, we view the beta weight family \citep{matsouaka2024causal} as defining a structured estimand trajectory anchored at the canonical ATE. When the indexing hyperparameter $\beta$ equals zero, the beta weights reduce to uniform weights and the corresponding WATE is exactly the ATE. As $\beta$ increases, observations with extreme propensity scores receive progressively less weight, and the resulting WATE estimators become more stable under limited overlap. Viewed from this perspective, $\beta$ acts as an estimand-level regularization parameter that traces the trade-off between statistical stability and departure from the canonical ATE. Under suitable conditions, this estimand trajectory admits a polynomial approximation around $\beta=0$. We exploit this structure by estimating a sequence of WATEs at positive values of $\beta$, fitting a polynomial to these estimates, and extrapolating the fitted trajectory to $\beta=0$. We refer to the resulting approach Polynomial approximation and Extrapolation to the Target estimand (PET). Importantly, PET treats the WATEs as intermediate quantities from which to recover the ATE, rather than as alternative final targets.

Provided that the approximation bias from the polynomial remainder is asymptotically negligible, the PET estimator is consistent and asymptotically normal for the ATE. Because its asymptotic variance is determined by the joint behavior of the more stable WATE estimators, PET can remain asymptotically normal under weaker overlap conditions, including settings in which the standard IPW estimator is unstable or has an infinite asymptotic variance. We characterize the approximation bias and propose a hyperparameter selection procedure that navigates the resulting bias-variance trade-off. We also develop an AIPW version of PET that, subject to the same approximation condition, is consistent if either the propensity score model or the outcome model is correctly specified. The key distinction from existing weighting and trimming approaches is that PET retains the canonical ATE as the prespecified target while using multiple modified estimands to inform its estimation. Moreover, whereas \citet{ma2020robust} and \citet{chaudhuri2025heavy} recover the ATE through trimming and tail-specific bias correction, PET extrapolates across a low-dimensional trajectory of systematically related causal estimands. Thus, PET does not eliminate extrapolation, but places it on a structured estimand scale directly anchored at the canonical ATE. To our knowledge, this use of multiple WATEs as intermediate quantities for recovering the ATE has not previously been considered.

The remainder of this article is organized as follows. In Section \ref{sec:prelim}, we review the estimands, standard IPW estimator, and related assumptions. %We then discuss the relationship between the overlap assumption and the existence of the variance of IPW estimators.
In Section \ref{sec:ATE-PET}, we introduce the polynomial representations of causal estimands defined by the beta weight family. Based on this insight, we formally describe the PET method and study its theoretical properties. In Section \ref{sec:app-bias}, we provide practical discussion of the PET method to assist with practice. Section \ref{sec:simu} includes simulation studies to compare the performance of the proposed method with existing approaches and demonstrate the merits of PET method under limited overlap. Section \ref{sec:appl} provides two complementary real data applications to illustrate how PET method can be used in practice.

\section{Preliminaries}\label{sec:prelim}
We consider a causal study with $n$ units. Let $A\in\{0,\, 1\}$, $X\in\mathcal{X}\subset\mathbb{R}^{p}$ and $(Y^{1},\, Y^{0})\in \mathbb{R}^{2}$ denote the treatment, a vector of covariates measured prior to treatment, and potential outcomes, respectively.
Additionally, under the Stable Unit Treatment Value Assumption \citep{rosenbaum1983central}, the observed outcome is represented as $Y=AY^{1}+(1-A)Y^{0}$.
Suppose $n$ independent and identically distributed copies of $(A,X,Y)$ are observed and denoted by $(A_{i},X_{i},Y_{i})$. Under the potential outcomes framework, a canonical target of inference is the ATE, denoted by $\tau={\rm E}[Y^{1}-Y^{0}]$. Assuming conditional exchangeability $A\indep(Y^{1},\, Y^{0})\mid X$ and positivity $0<e(X)<1$, the ATE is identifiable using the device of propensity score $e_{i}=e(X_{i})={\rm Pr}(A=1\mid X_{i})$ \citep{rosenbaum1983central}.
In the following discussions, we assume that the potential outcomes have finite second moments:\ ${\rm E}\left[\left(Y^a\right)^2\right]<\infty$ ($a=0,1$).
To simplify the presentation of the main idea, we also assume that the propensity score is known in this section.

\subsection{Inverse probability weighting}
%In this manuscript, we mainly consider the inverse probability weighting (IPW) estimator. 
The standard, H\'{a}jek-type IPW estimator for the ATE is written as
$$
\hat{\tau}_{\text{IPW}}
=
\frac{\sum_{i=1}^{n}{A_iY_i}/{e(X_i)}}{\sum_{i=1}^{n}{A_i}/{e(X_i)}}
-
\frac{\sum_{i=1}^{n}{(1-A_i)Y_i}/\{1-e(X_i)\}}{\sum_{i=1}^{n}{(1-A_i)}/\{1-e(X_i)\}}.
$$
It is well known that the IPW estimator has consistency and asymptotic normality under some regularity conditions.
One of these conditions is known as the {\it strict overlap assumption} \citep{d2021overlap}:
\begin{assu}(Strict overlap) \label{assu1}
There exists $\delta\in(0,1)$ such that for all $x\in\mathcal{X}$, $\delta\leq e(x)\leq1-\delta$.
\end{assu}
Under Assumption \ref{assu1}, the IPW estimator satisfies
$\sqrt{n}(\hat{\tau}_{\text{IPW}}-\tau)
\stackrel{d}{\to}
N(0,\sigma_{\text{IPW}}^2)$,
where the asymptotic variance takes the form of
$
\sigma_{\text{IPW}}^2={\rm E}\left[{\left(Y^1-\theta_1^{0}\right)^2}/{e(X)}\right]+{\rm E}\left[{\left(Y^0-\theta_0^{0}\right)^2}/\{1-e(X)\}\right]$. Here, $\theta_a^{0}={\rm E}[Y^a]$.
In fact, for the first component of $\sigma_{\text{IPW}}^2$, we have 
$
{\rm E}\left[{\left(Y^1-\theta_1^{0}\right)^2}/{e(X)}\right]
\leq
{{\rm E}\left[\left(Y^1-\theta_1^{0}\right)^2\right]}/{\delta}
<
\infty.
$
The second component can be handled in the same manner.
Therefore, the strict overlap assumption is a sufficient condition for the existence of $\sigma_{\text{IPW}}^2$. However, it is known that when Assumption \ref{assu1} is violated or nearly violated, the IPW estimator $\hat{\tau}_{\text{IPW}}$ has an excessively large variance or does not converge to a normal distribution \citep{li2019addressing,heiler2021valid}.
A design-based option to overcome this problem is modifying the target estimand away from the canonical ATE.

\subsection{Targeting a class of weighted average treatment effects}
Following \citet{hirano2003efficient}, the class of weighted average treatment effect (WATE) is
$$
\tau_w
=
\frac{{\rm E}[w(X)\tau(X)]}{{\rm E}[w(X)]},
$$
where $\tau(X)={\rm E}[Y^{1}-Y^{0}\mid X]$ is the conditional average treatment effect.
Hereinafter, we assume $\sup_{x\in\mathcal{X}}|\tau(x)|<\infty$. It is easy to see that when $w(x)\equiv1$, $\tau_w=\tau$ becomes the canonical ATE. The WATE includes many types of estimands such as the trimmed ATE with $w(x)=1\{\alpha<e(x)<1-\alpha\}$ with $\alpha\in(0,0.5)$ \citep{crump2009dealing}, the ATO with overlap weights $w(x)=e(x)(1-e(x))$, and more recently, a restricted class of WATE with
$$
w(x)
=
w_\beta(x)
=
\{e(x)(1-e(x))\}^{\beta},
\ \ \ (\beta\geq0)
$$
referred to as the beta weight family \citep{matsouaka2024causal}. From here, we focus on the beta weight family (hereafter simply referred to as WATE) with $\beta\in[0,1)$ and define the WATE as $\tau_\beta$. Here, note that ${\rm E}[w_{\beta}(X)]>0$ from the positivity assumption, and $\tau_0=\tau$.

Focusing on the beta weight family, the weighting estimator is given by
\begin{align}
\hat{\tau}_\beta
&=
\frac{\sum_{i=1}^{n}{A_i}w_\beta(X_i)Y_i/{e(X_i)}}{\sum_{i=1}^{n}{A_i}w_\beta(X_i)/{e(X_i)}}
-
\frac{\sum_{i=1}^{n}(1-A_i)w_\beta(X_i)Y_i/\{1-e(X_i)\}}{\sum_{i=1}^{n}{(1-A_i)}w_\beta(X_i)/\{1-e(X_i)\}} %\\
% &=
% \frac{1}{n}\sum_{i=1}^{n}\left[\frac{A_i}{e(X_i)}\left(\frac{w_\beta(X_i)}{n^{-1}\sum_{j=1}^{n}\frac{A_j}{e(X_j)}w_\beta(X_j)}\right)
% -
% \frac{1-A_i}{1-e(X_i)}\left(\frac{w_\beta(X_i)}{n^{-1}\sum_{j=1}^{n}\frac{1-A_j}{1-e(X_j)}w_\beta(X_j)}\right)\right]Y_i, 
\label{WATE_IPW}
\end{align}
and the asymptotic variance $\text{Var}(\hat{\tau}_{\beta})$ is
\begin{align*}%\label{var_WATE}
\sigma_{\beta}^2
% &=
% Var(\hat{\tau}_{\beta})=
% \frac{{\rm E}\left[\frac{w_{\beta}(X)^2}{e(X)}\left(Y^1-\theta_{1\beta}^{0}\right)^2\right]}{{\rm E}[w_{\beta}(X)]^2}
% +\frac{{\rm E}\left[\frac{w_{\beta}(X)^2}{1-e(X)}\left(Y^0-\theta_{0\beta}^{0}\right)^2\right]}{{\rm E}[w_{\beta}(X)]^2} \nonumber \\
=\frac{{\rm E}\left[e(X)^{2\beta-1}(1-e(X))^{2\beta}\left(Y^1-\theta_{1\beta}^{0}\right)^2\right]
+{\rm E}\left[e(X)^{2\beta}(1-e(X))^{2\beta-1}\left(Y^0-\theta_{0\beta}^{0}\right)^2\right]}{{\rm E}[w_{\beta}(X)]^2},
\end{align*}
where $\theta_{a\beta}^{0}={\rm E}[w_\beta(X)Y^a]/{\rm E}[w_\beta(X)]$. From the form of $\sigma_{\beta}^2$, neither $e(X)^{-1}$ nor $(1-e(X))^{-1}$ appears when $\beta\geq0.5$. 
%In other words, the strict overlap assumption is not required for $\sigma_{\beta}^2$ to be finite.
When $0<\beta<0.5$, negative powers of $e(X)$ and $1-e(X)$ remain in $\sigma_{\beta}^2$, but their exponents $2\beta-1$ are closer to zero than the exponents $-1$ appearing in $\sigma_{\text{IPW}}^2$.
Consequently, the inflation in the variance caused by extreme propensity scores is less severe than the standard IPW estimator.
On the other hand, since $e(X)^{2\beta}<1$, $(1-e(X))^{2\beta}<1$, $e(X)^{2\beta-1}< e(X)^{-1}$, and $(1-e(X))^{2\beta-1}<(1-e(X))^{-1}$, one can show
\begin{align}
\sigma_{\beta}^2&
<
\frac{1}{{\rm E}[w_{\beta}(X)]^2}
\left\{{\rm E}\left[\frac{\left(Y^1-\theta_{1\beta}^{0}\right)^2}{e(X)}\right]
+{\rm E}\left[\frac{\left(Y^0-\theta_{0\beta}^{0}\right)^2}{1-e(X)}\right]\right\}. \label{ineq_var}
\end{align}
Thus, the strict overlap assumption is a sufficient condition for the upper bound in \eqref{ineq_var} to be finite, but not necessarily required for $\sigma_{\beta}^2$ itself to be finite. Finally, when $\beta=0$, it is immediate that $\sigma_{\beta}^2=\sigma_{\text{IPW}}^2$.

The above discussion suggests that the asymptotic variance of the weighting estimator for the WATE, $\hat{\tau}_\beta$, may be smaller than that of the standard IPW estimator, but still remain bounded under weaker overlap conditions rather than the strict overlap assumption. In this sense, $\hat{\tau}_\beta$ is expected to be more stable than $\hat{\tau}_{\text{IPW}}$ under limited overlap. This gain in stability, however, comes at the cost of changing the target estimand. That is, the WATE estimand generally differs from the canonical ATE estimand, and its interpretation depends on the choice of weighting function and may be less straightforward \citep{rizk2025and}. This trade-off motivates an approach that retains the ATE as the target while exploiting the greater stability of WATE estimators under limited overlap.
%However, the WATE estimand is generally different from the canonical ATE estimand, and depending on the choice of the weight, the interpretation of the WATE may be less straightforward than that of the ATE \citep{rizk2025and}. Therefore, it is desirable to develop an estimator of the ATE that is more precise than the standard IPW estimator while reducing the bias relative to the ATE that arises from using the IPW estimator for the WATE.

\section{Polynomial approximation and extrapolation}\label{sec:ATE-PET}
\subsection{Polynomial approximation for the WATE}\label{sec:poly}
%The ATE $\tau$ is generally different from the WATE $\tau_\beta$.However, we 
Although the WATE estimands are generally different from the canonical ATE estimand, we first show that there is a polynomial relationship connecting these estimands. To establish this property, we first consider a Taylor expansion of $h_\beta(x):=w_{\beta}(x)/{\rm E}[w_{\beta}(X)]$:
$$
h_{\beta}(x)=\sum_{\ell=0}^{q}\frac{h_{\beta}^{(\ell)}(x)|_{\beta=0}}{\ell!}\beta^{\ell}+\frac{h_{\beta}^{(q+1)}(x)|_{\beta=c(x)}}{(q+1)!}\beta^{q+1}\ \ \ (c(x)\in(0,\beta)).
$$
Here, $h_{\beta}^{(\ell)}(x)$ denotes the $\ell$-th derivative of $h_\beta(x)$ with respect to $\beta$.
Based on this expansion, the following proposition holds (proof in Appendix \ref{appa}).
\begin{prp}\label{prop2}
Under a regularity condition described in Appendix \ref{appa},
%a sufficient condition for the following relationship to hold
the following relationship holds:
\begin{align}
\tau_{\beta}=\tau+\sum_{\ell=1}^{q}\frac{1}{\ell!}{\rm E}\left[\left.h_{\beta}^{(\ell)}(X)\right|_{\beta=0}\tau(X)\right]\beta^{\ell}+\frac{1}{(q+1)!}{\rm E}\left[\left.h_{\beta}^{(q+1)}(X)\right|_{\beta=c(X)}\tau(X)\right]\beta^{q+1}. \label{poly_inf}
\end{align}
%is that, for all $r=0,1,\dots,q+1$, there exists $C_r<\infty$ such that, for all $x\in\mathcal{X}$,
%$$
%\sup_{\eta\in[0,\beta]}\left|h_{\eta}^{(r)}(x)\right|\leq C_r\left(1+|\log \{e(x)(1-e(x))\}|^r\right).
%$$
\end{prp}
%The proof is given in Appendix \ref{appa}.

Proposition \ref{prop2} shows that the mapping $\beta\mapsto\tau_\beta$ admits a polynomial expansion around $\beta=0$, whose intercept is the canonical ATE. Consequently, if the WATEs were available at several positive values of $\beta$, one could in principle recover $\tau$ by extrapolating this relationship to $\beta=0$. This observation motivates estimating a collection of WATEs at values $\beta\in D$, for which the corresponding estimators have finite asymptotic variances, and then using polynomial extrapolation to estimate the ATE. The accuracy of this approximation is governed by the remainder term in \eqref{poly_inf}. Hereinafter, we define $D\subset[0,1)$ as the set of $\beta$ for which $\sigma_\beta^2$ is finite, and assume that $D\neq\varnothing$.
%Then, for $\beta\in D$, $\hat{\tau}_\beta$ has the asymptotic variance $\sigma_\beta^2$.
For $\beta_k\in D$ ($\beta_1<\cdots<\beta_K$), we further define the vectors of the the WATE estimators and their true estimand values as
$\hat{\bld{\tau}}=(\hat{\tau}_{\beta_1},\dots,\hat{\tau}_{\beta_K})^{\top}$ and 
$\bld{\tau}
=
(\tau_{\beta_1},\dots,\tau_{\beta_K})^{\top}.$ Additionally, define $v_k={(1,\beta_k,\dots,\beta_k^{q})}^{\top}$ and
$$
V
=
(v_1,\dots,v_K)
=
\left(
\begin{array}{ccc}
1&\cdots&1\\
\beta_1&\cdots&\beta_K\\
\vdots&&\vdots\\
\beta_1^{q}&\cdots&\beta_K^{q}
\end{array}
\right)
=
\left(
\begin{array}{c}
\bld{1}^{\top}_{K}\\
B
\end{array}
\right),
$$
where $q\in\mathbb{N}$ ($q\leq K-1$) and $VV^{\top}$ is nonsingular. Writing $\bld{\gamma}={(\gamma_0,\bld{\gamma}_1^{\top})}^{\top}$, we consider the following polynomial regression model:
\begin{align}
\hat{\tau}_{\beta_k}
=
v_k^{\top}\bld{\gamma}+\epsilon_k
=
\gamma_0+\sum_{\ell=1}^{q}\gamma_{\ell}\beta^{\ell}_k+\epsilon_k. \label{mod1}
\end{align}
Using the defined vector and matrix, we can write $\hat{\bld{\tau}}
=
V^{\top}\bld{\gamma}+\bld{\epsilon}
=
\gamma_0\bld{1}_{K}+B^{\top}\bld{\gamma}_1+\bld{\epsilon}$, where $\bld{\epsilon}=(\epsilon_1,\dots,\epsilon_K)^{\top}$.
The polynomial regression model \eqref{mod1} can be regarded as a $q$-dimensional approximation to \eqref{poly_inf}.

\subsection{Proposed estimator for the average treatment effect}
\begin{figure}[ht!]
\begin{center}
\begin{tabular}{c}
\includegraphics[width=11cm]{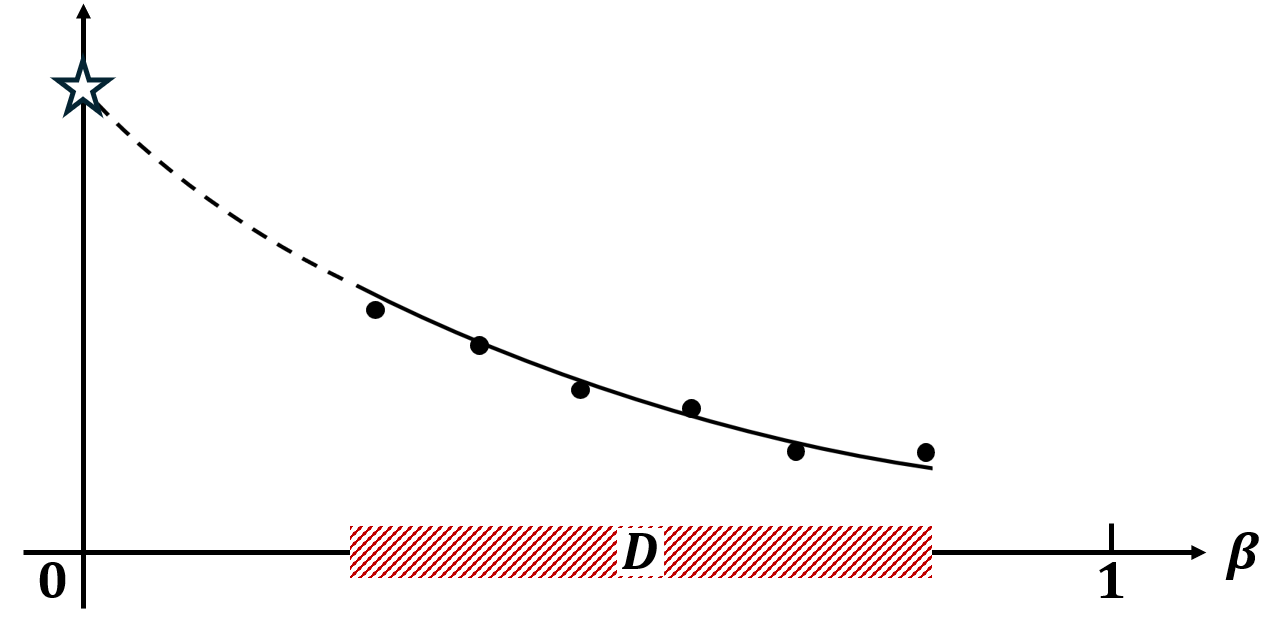}
\end{tabular}\caption{Conceptual illustration of the PET method. The red shaded area corresponds to $D \subset [0,1)$. The procedure consists of three steps:\ 1) implementing estimators for WATE (black dots), 2) modeling these estimates using a polynomial function (solid line), and 3) extrapolating to recover the canonical ATE (dashed line). The star indicates the resulting proposed ATE estimator.}
\label{fig0}
\end{center}
\end{figure}

Under model \eqref{mod1}, to estimate $\bld{\gamma}$, we consider the estimating equation $V(\hat{\bld{\tau}}-V^{\top}\bld{\gamma})=0$, whose solution is $\hat{\bld{\gamma}}={\left(
\hat{\gamma}_0,
\hat{\bld{\gamma}}_1^{\top}
\right)}^{\top}
$.
% where $\bar{P}_B=I_K-P_B=I_K-B^{\top}\left(BB^{\top}\right)^{-1}B$. 
Our proposed estimator, referred to as the Polynomial approximation and Extrapolation to the Target estimand (PET), is formally given by
\begin{align}
\hat{\tau}_{P}
&=\hat{\gamma}_0
=
{\bld{1}_K^{\top}\bar{P}_B\hat{\bld{\tau}}}/\left(\bld{1}_K^{\top}\bar{P}_B\bld{1}_K\right) \nonumber \\
&=
\sum_{i=1}^{n}
\sum_{k=1}^{K}\alpha_k\left(\frac{A_iw_{\beta_k}(X_i)/e(X_i)}{\sum_{j=1}^{n}{A_j}w_{\beta_k}(X_j)/e(X_j)}-\frac{(1-A_i)w_{\beta_k}(X_i)/\{1-e(X_i)\}}{\sum_{j=1}^{n}{(1-A_j})w_{\beta_k}(X_j)/\{1-e(X_j)\}}\right)Y_i, \label{PET1}
\end{align}
where $\bar{P}_B=I_K-P_B=I_K-B^{\top}\left(BB^{\top}\right)^{-1}B$, $I_K$ is the $K$-dimensional identity matrix, 
$
\alpha_k
=\left\{1-\sum_{k'=1}^{K}b_{k'k}\right\}/\left(\bld{1}_K^{\top}\bar{P}_B\bld{1}_K\right),
$ and $b_{k'k}$ denotes the $(k',k)$ component of $P_B$. 
This procedure can be regarded as a polynomial approximation based on model \eqref{mod1}, followed by extrapolation to $\beta=0$. Figure \ref{fig0} provides a conceptual illustration.
 From another perspective, compared with \eqref{WATE_IPW}, the proposed estimator \eqref{PET1} can be viewed as a weighted combination of WATE estimators, where the $k$-th estimator is assigned weight $\alpha_k$.

We prove the following result stating the asymptotic property of the PET estimator,
\begin{thm}\label{theo1}
Assume that the propensity score is known and
\begin{align}
\bar{P}_B(\bld{\tau}
-\tau\bld{1}_K)=\bld{0}_K. \label{assu:poly}
\end{align}
Then, $\sqrt{n}\left(\hat{\tau}_{P}-\tau\right)\stackrel{d}{\to}N(0,\sigma^2)$, where
\begin{align}
\sigma^2
&=
\sum_{k=1}^{K}\frac{\alpha_k^2}{{\rm E}\left[w_{\beta_{k}}(X)\right]^2}\left\{{\rm E}\left[\frac{w_{\beta_{k}}(X)^2(Y^1-\theta^0_{1k})^2}{e(X)}\right]+{\rm E}\left[\frac{w_{\beta_{k}}(X)^2(Y^0-\theta^0_{0k})^2}{1-e(X)}\right]\right\}\nonumber\\
&\hspace{0.5cm}+\sum_{k=1}^{K}\sum_{k'\neq k}\frac{\alpha_k\alpha_{k'}}{{\rm E}\left[w_{\beta_{k}}(X)\right]{\rm E}\left[w_{\beta_{k'}}(X)\right]}\left\{{\rm E}\left[\frac{w_{\beta_{k}}(X)w_{\beta_{k'}}(X)(Y^1-\theta^0_{1k})(Y^1-\theta^0_{1k'})}{e(X)}\right]\right.\nonumber\\
&\hspace{1cm}\left.+{\rm E}\left[\frac{w_{\beta_{k}}(X)w_{\beta_{k'}}(X)(Y^0-\theta^0_{0k})(Y^0-\theta^0_{0k'})}{1-e(X)}\right]\right\} \nonumber \\
&=\sum_{k=1}^{K}\alpha_k^2\text{Var}(\hat{\tau}_{\beta_k})+\sum_{k=1}^{K}\sum_{k'\neq k}\alpha_k\alpha_{k'}\text{Cov}(\hat{\tau}_{\beta_k},\hat{\tau}_{\beta_{k'}}). \label{avar}
\end{align}
Here, $\theta^0_{ak}=\theta^0_{a\beta_k}$ and
$$
\text{Cov}(\hat{\tau}_{\beta},\hat{\tau}_{\beta'})=\frac{{\rm E}\left[\displaystyle\frac{w_{\beta}(X)w_{\beta'}(X)(Y^1-\theta^0_{1\beta})(Y^1-\theta^0_{1\beta'})}{e(X)}\right]+{\rm E}\left[\displaystyle\frac{w_{\beta}(X)w_{\beta'}(X)(Y^0-\theta^0_{0\beta})(Y^0-\theta^0_{0\beta'})}{1-e(X)}\right]}{{\rm E}\left[w_{\beta}(X)\right]{\rm E}\left[w_{\beta'}(X)\right]}.
$$
\end{thm}
%The proof is given in Appendix \ref{appa}.Note that $\sigma_\beta^2=Var(\hat{\tau}_{\beta_k})$ exists if $\beta\in D$.
Theorem \ref{theo1} shows that the asymptotic variance of $\hat{\tau}_P$ can be represented as a weighted sum of the variances of $\hat{\tau}_{\beta_k}$ and the covariances between $\hat{\tau}_{\beta_k}$ and $\hat{\tau}_{\beta_{k'}}$; 
thus, its variance does not directly depend on that of $\hat{\tau}_{\text{IPW}}$.
In this sense, the proposed estimator remains valid under weaker overlap conditions instead of Assumption \ref{assu1}, and potentially have a smaller variance than $\hat{\tau}_{\text{IPW}}$ under limited overlap without compromising consistency.
\begin{rem}
Assumption \eqref{assu:poly} implies that the approximation bias due to the polynomial approximation can be ignored.
In other words, the remainder beyond the $q$-th order polynomial in \eqref{poly_inf} is assumed to be ignorable.
This assumption is fundamental to deriving the asymptotic result, but it is only a sufficient condition.
We further discuss the validity of this assumption and the possibility of relaxing it in Section \ref{sec:appr}.
\end{rem}

In practice, the propensity score is typically unknown and must be estimated. Hereinafter, we follow the practical convention and assume that the propensity score is modeled by logistic regression, 
$
e_i(\zeta)
=
{\exp\{\varphi(x_i;\zeta)\}}/{\left[1+\exp\{\varphi(x_i;\zeta)\}\right]},$ 
and $\zeta$ is estimated as the maximum likelihood estimator, denoted as $\hat{\zeta}$. Under the setting, the asymptotic normality still holds, as outlined in Theorem \ref{coro1}.
\begin{thm}\label{coro1}
Assume that \eqref{assu:poly} holds.
When the estimated propensity score model $e_i(\hat{\zeta})$ is plugged into $\hat{\tau}_{\beta_{k}}$, we have $\sqrt{n}\left(\hat{\tau}_P-\tau\right)\stackrel{d}{\to}N(0,{\sigma'}^2)$, where 
\begin{align}
{\sigma'}^2
=
\sigma^2
+
\Lambda.\label{avar2}
\end{align}
\end{thm}
As discussed in Appendix \ref{appa}, the additional term due to estimating the propensity score $\Lambda$ can take both positive and negative values. Hence, a variance estimator for $\hat{\tau}_P$ based on \eqref{avar} may overestimate or underestimate the true variance.
The sign of $\Lambda$ is determined by the covariance between the weight $w_\beta$ and the outcome $Y$ \citep{orihara2025re}.
%More details can be found in the influence function described in Appendix \ref{appa}.
% In the following simulation experiment, we use the plug-in estimator of the asymptotic variance \eqref{avar2}.

\subsection{Augmentation and double robustness}\label{sec:aug}
The PET estimator developed above addresses instability due to limited overlap by extrapolating from a collection of WATE estimators whose asymptotic variances are finite. However, because its constituent estimators rely only on weighting, PET does not exploit outcome information and may remain sensitive to misspecification of the propensity score model. We therefore develop an augmented version of PET. The key observation is that PET is a linear combination of WATE estimators with coefficients $\alpha_k$ that do not depend on the observed data. Consequently, augmentation can be applied to each component WATE estimator and then carried through the same polynomial extrapolation. The two components therefore play complementary roles. The choice of $\beta_k\in D$ and subsequent extrapolation mitigate instability under limited overlap, whereas outcome regression can improve efficiency and provide additional robustness to nuisance model misspecification.

To formalize this construction, under \eqref{assu:poly} and a known propensity score, the influence function of the PET estimator is
$$
\phi_i
=
\sum_{k=1}^{K}\alpha_k\left[\frac{W_{ki}A_i}{{\rm E}[w_{\beta_k}(X)]}(Y_i-\theta^{0}_{1k})-\frac{W_{ki}(1-A_i)}{{\rm E}[w_{\beta_k}(X)]}(Y_i-\theta^{0}_{0k})\right],
$$
where
$
W_{ki}={w_{\beta_k}(X_i)}/\left\{A_ie(X_i)+(1-A_i)(1-e(X_i))\right\}$. 
Since the influence function has the same form as the weighting estimator for the WATE \citep{mao2019propensity}, except for $\alpha_k$ which does not depend on the observed data, we consider the augmentation space consisting of functions of the form $(A-e(X))L(X)$ ($L(X)$ is an arbitrary function).
Specifically,
\begin{align}
(A-e(X))L(X)
%&=-(A-e(X))\sum_{k=1}^{K}\alpha_k\left[\frac{w_{\beta_k}(X)}{{\rm E}[w_{\beta_k}(X)]}\left(\frac{m_1(X)-\theta^{0}_{1k}}{e(X)}+\frac{m_0(X)-\theta^{0}_{0k}}{1-e(X)}\right)\right] \nonumber \\
&=
\sum_{k=1}^{K}\alpha_k\left[\frac{w_{\beta_k}(X)}{{\rm E}[w_{\beta_k}(X)]}\left\{\left(1-\frac{A}{e(X)}\right)\left(m_1(X)-\theta^{0}_{1k}\right)\right.\right. \nonumber \\ 
&\hspace{2cm}
\left.\left.-\left(1-\frac{1-A}{1-e(X)}\right)\left(m_0(X)-\theta^{0}_{0k}\right)\right\}\right]. \label{aug1}
\end{align}
Here, $m_a(X)={\rm E}[Y\mid A=a,X]$ is the outcome regression function and the outcome model is denoted by $m_a(X;\xi)$.
Using the augmentation term \eqref{aug1}, we then obtain the following influence function:
\begin{align*}
\phi_{\text{AIF},\, i}
&=
\phi_{i}+(A_i-e(X_i))L(X_i) \\ 
% &=
% \sum_{k=1}^{K}\alpha_k\left[\frac{W_{ki}A_i}{{\rm E}[w_{\beta_k}(X)]}(Y_i-\theta^{0}_{1k})-\frac{W_{ki}(1-A_i)}{{\rm E}[w_{\beta_k}(X)]}(Y_i-\theta^{0}_{0k}) \right. \\ 
% &\hspace{0.5cm}
% \left. +\frac{w_{\beta_k}(X_i)}{{\rm E}[w_{\beta_k}(X)]}\left\{\left(1-\frac{A}{e(X)}\right)\left(m_1(X)-\theta^{0}_{1k}\right)-\left(1-\frac{1-A}{1-e(X)}\right)\left(m_0(X)-\theta^{0}_{0k}\right)\right\}\right] \\
&=\sum_{k=1}^{K}\alpha_k\frac{w_{\beta_k}(X_i)}{{\rm E}[w_{\beta_k}(X)]}\left[\left(\frac{A_i}{e(X_i)}-\frac{1-A_i}{1-e(X_i)}\right)Y_i-\left(\theta^{0}_{1k}-\theta^{0}_{0k}\right) \right. \\ 
&\hspace{0.5cm}
\left. +\left\{\left(1-\frac{A_i}{e(X_i)}\right)m_1(X_i)-\left(1-\frac{1-A_i}{1-e(X_i)}\right)m_0(X_i)\right\}\right].
\end{align*}
The final expression shows that $\phi_{\text{AIF},i}$ is the same linear combination of the augmented influence functions for the $K$ constituent WATE estimators. Thus, augmentation and extrapolation can be performed componentwise. Each WATE estimator is first augmented using the corresponding outcome regression, and the resulting estimators are then combined using the original extrapolation coefficients $\alpha_k$. Here, since $\tau_{\beta_k}=\theta^{0}_{1k}-\theta^{0}_{0k}$ and
$\sum_{k=1}^{K}\alpha_k\tau_{\beta_k}=\tau$ 
from \eqref{assu:poly} (see also Appendix \ref{appa}), the augmented PET (AIPW-PET) estimator is given by
\begin{align*}
\hat{\tau}_{P}^{\text{DR}}
&=
\frac{1}{n}\sum_{i=1}^{n}\sum_{k=1}^{K}\alpha_k\frac{w_{\beta_k}(X_i;\hat{\zeta})}{n^{-1}\sum_{j=1}^{n}w_{\beta_k}(X_j;\hat{\zeta})}\left[\left(\frac{A_i}{e_i(\hat{\zeta})}-\frac{1-A_i}{1-e_i(\hat{\zeta})}\right)Y_i\right. \\
&\hspace{1.5cm}\left.
+\left\{\left(1-\frac{A_i}{e_i(\hat{\zeta})}\right)m_1(X_i;\hat{\xi})-\left(1-\frac{1-A_i}{1-e_i(\hat{\zeta})}\right)m_0(X_i;\hat{\xi})\right\}\right],
\end{align*}
and its variance estimator is based on
\begin{align}
\sigma^2_{DR}={\rm E}\left[\phi_{\text{AIF}}^2\right]. \label{avar3}
\end{align}
When both nuisance models are correctly specified, incorporating prognostic outcome information can improve efficiency relative to the IPW-PET estimator. Beyond this potential efficiency gain, augmentation provides robustness to misspecification of either nuisance model. That is, when the propensity score model is correctly specified, the proposed semiparametric estimator is consistent by construction. When the outcome model is correctly specified (i.e., $m_a(X;\xi^{0})\equiv m_a(X)$), 
\begin{align*}
\phi_{\text{AIF},\, i}
&=
\sum_{k=1}^{K}\alpha_k\left[\frac{w_{\beta_k}(X_i)}{{\rm E}[w_{\beta_k}(X)]}\left(m_1(X_i)-m_0(X_i)-\theta^{0}_{1k}+\theta^{0}_{0k}\right) \right. \\
&\hspace{0.5cm}
\left. +\frac{w_{\beta_k}(X_i)}{{\rm E}[w_{\beta_k}(X)]}\left(\frac{A_i}{e(X_i)}(Y_i-m_1(X_i))-\frac{1-A_i}{1-e(X_i)}(Y_i-m_0(X_i))\right)\right],
\end{align*}
and the expectation of the latter part is zero.
The important point is that there is no explicit term involving the inverse of the propensity score in the former part.
Summarizing the above discussion, we obtain the following theorem.
\begin{thm}\label{theo2}
Assume that \eqref{assu:poly} holds for both correctly specified and misspecified propensity score models.
If either the propensity score model is correctly specified ($e_i(\zeta^{0})\equiv e(X_i)$), or the outcome model is correctly specified ($m_a(X_i;\xi^{0})\equiv m_a(X_i)$), then $\hat{\tau}_{P}^{DR}\stackrel{P}{\to}\tau$.
\end{thm}

\begin{rem}\label{rmk2}
Although Theorem \ref{theo2} establishes double robustness with respect to the nuisance models, validity for the ATE remains conditional on the polynomial extrapolation condition. Thus, double robustness does not protect against approximation bias arising when this condition fails. This distinguishes AIPW-PET from the standard AIPW estimator, which targets the ATE directly and does not require polynomial extrapolation. Accordingly, the influence function of AIPW-PET derived under \eqref{assu:poly} generally differs from the efficient influence function underlying the standard AIPW estimator \citep{tsiatis2006semiparametric}. By avoiding direct estimation at $\beta=0$, AIPW-PET may offer greater stability under limited overlap, at the cost of potential approximation bias when the polynomial relationship holds only approximately. We study this approximation bias in Section \ref{sec:appr} and compare the finite-sample performance of the two estimators in the simulation experiments.
%The target estimand of the PET methods is $\tau$ (the canonical ATE). However, the theoretical validity of the PET estimators depends on Assumption \eqref{assu:poly}. Accordingly, the derived influence function $\phi_{\text{AIF},\, i}$ under this assumption differs from the standard AIPW influence function for $\tau$ \citep{tsiatis2006semiparametric}. As discussed in Section \ref{sec:appr}, the approximation bias (the left-hand side of \eqref{assu:poly}) is expected to be small, at least asymptotically. Consequently, the proposed AIPW-PET estimator is expected to exhibit different behavior from the standard AIPW estimator, especially in finite-sample settings. This difference is examined in the ensuing simulation experiments.
\end{rem}

\section{Practical considerations for implementation}\label{sec:app-bias}
\subsection{Approximation bias}\label{sec:appr}
In Theorem \ref{theo1}, \ref{coro1}, and \ref{theo2}, we assume \eqref{assu:poly}, which implies that the approximation bias due to the polynomial approximation \eqref{mod1} can be ignored.
To evaluate this, let us consider the deviation estimator $\hat{\tau}_P-\tau$. We define
$
\gamma_{0}^{0}
={\bld{1}_K^{\top}\bar{P}_B\bld{\tau}}/\left(\bld{1}_K^{\top}\bar{P}_B\bld{1}_K\right)$. Then, we have 
\begin{align*}
\hat{\tau}_P-\tau
% &=
% \hat{\tau}_P-\gamma_{0}^{0}+\gamma_{0}^{0}-\tau \\
&=
\gamma_{0}^{0}-\tau+\frac{\bld{1}_K^{\top}\bar{P}_B}{\bld{1}_K^{\top}\bar{P}_B\bld{1}_K}(\hat{\bld{\tau}}-\bld{\tau}) =
\gamma_{0}^{0}-\tau+O_{p}\left(\frac{1}{\sqrt{n}}\right).
\end{align*}
By Taylor series and under the condition in Proposition \ref{prop2},
\begin{align}
\gamma_{0}^{0}-\tau
&=
\frac{\bld{1}_K^{\top}\bar{P}_B}{\bld{1}_K^{\top}\bar{P}_B\bld{1}_K}\left(\tau\bld{1}_K+B^{\top}\bld{\gamma}+\left(
\gamma_{k,q+1}\beta_k^{q+1}
\right)_{k=1,\dots,K}\right)-\tau \nonumber \\
&=
\frac{\bld{1}_K^{\top}\bar{P}_B}{\bld{1}_K^{\top}\bar{P}_B\bld{1}_K}\left(
\gamma_{k,q+1}\beta_k^{q+1}
\right)_{k=1,\dots,K}. \label{apperr1}
\end{align}
Here, the expansion coefficient is given by
$$
\gamma_{k,q+1}
=
\frac{{\rm E}\left[\left.h_{\beta}^{(q+1)}(X)\right|_{\beta=c_k(X)}\tau(X)\right]}{(q+1)!}
\ \ \ (c_k(x)\in (0,\beta_k)).
$$
Therefore, condition \eqref{assu:poly} implies that the right-hand side of \eqref{apperr1}, the remainder term in \eqref{poly_inf}, can be ignored.
In this sense, we assume that there is no approximation bias in deriving our key results. Although this condition does not hold theoretically in general, the approximation bias \eqref{apperr1} is expected to be small in many settings for the following reasons.
If $\gamma_{k,q+1}\equiv\gamma_{q+1}$, \eqref{apperr1} becomes
$
\gamma_{q+1}\frac{\bld{1}_K^{\top}\bar{P}_B}{\bld{1}_K^{\top}\bar{P}_B\bld{1}_K}\bld{b}_{q+1}$. Here, $\bld{b}_{q}=\left(\beta_1^{q},\dots,\beta_K^{q}\right)^{\top}$.
It is well known that $\text{Corr}(\bld{b}_{q},\bld{b}_{q+1})$ is very large \citep{bradley1979correlation}, and hence the bias in \eqref{apperr1} becomes small since
$
B^{\top}
=
\left(
\bld{b}_{1},\dots,\bld{b}_{q}
\right)$ and 
$\bar{P}_B\bld{b}_{q+1}
\approx
0$. Additionally, when $q$ is relatively large, $\gamma_{k,q+1}\equiv\gamma_{q+1}\approx0$, and the bias also becomes small.
Therefore, if a large approximation bias is expected, $q$ should be set to a large value to offset this potential bias.

In fact, to prove Theorem \ref{theo1} and \ref{coro1}, the following condition is sufficient; \eqref{assu:poly} is not necessary:
\begin{align}
\gamma_{0}^{0}-\tau=o\left(\frac{1}{\sqrt{n}}\right). \label{assu:poly2}
\end{align}
Condition \eqref{assu:poly2} can be justified from several perspectives.
By simple calculation, we obtain the following upper bound for \eqref{apperr1}:
\begin{align}
|\gamma_{0}^{0}-\tau|
<
\frac{\sup_{x\in\mathcal{X}}\{|\tau(x)|\}K(1+K)}{\bld{1}_K^{\top}\bar{P}_B\bld{1}_K}C^{K}_{q+1}\beta_K^{q+1}, \label{apperr2}
\end{align}
where
$$
C^{K}_{q+1}
=
\sup_{k=1,\dots,K}\left\{\frac{{\rm E}\left[\left|\left.h_{\beta}^{(q+1)}(X)\right|_{\beta=c_k(X)}\right|\right]}{(q+1)!}\right\}.
$$
From \eqref{apperr2}, it follows that taking $\beta_{Kn}=O((nK)^{-1/q})$ is sufficient to ensure \eqref{assu:poly2} if $C^{K}_{q+1}< C_1$ and $\bld{1}_K^{\top}\bar{P}_B\bld{1}_K/K> C_2$ for some positive constants $C_1$ and $C_2$.
Here, the former condition requires the $(q+1)$-th order derivative term, normalized by $(q+1)!$, to remain uniformly bounded, whereas the latter condition means that $\bar{P}_B\bld{1}_K$ is not asymptotically orthogonal to $\bld{1}_K$.
Note that this discussion is based on $\beta_{Kn}\to0$; thus, it is necessary to replace the inverse term in $\bar{P}_B$ with a generalized inverse in the theoretical discussion.
Alternatively, for a fixed $\beta_K<1$, as a rough order argument based on the term $\beta_K^{q+1}$ in \eqref{apperr2}, $q_n\asymp O(\log(nK))$ provides a useful reference order when aiming for the approximation bias rate in \eqref{assu:poly2} (noting that $q\leq K-1$).
Further details are provided in Appendix \ref{app:app_bias}.

\subsection{Hyperparameter selection}\label{sec:hyp}
To implement the PET method, there are four hyperparameters to be specified before analysis:\ $K$, $\beta_1$, $\beta_K$, and $q$.
In practice, $K$ can be chosen sufficiently large, making the specific choice of $K$ less critical.
When $\beta_K$ is close to $0$, the variance becomes larger than when $\beta_K$ is close to $1$, as seen from the form of the asymptotic variance \eqref{avar}.
In contrast, both $C^{K}_{q+1}$ and $\beta_K^{q+1}$ in \eqref{apperr2} tend to be smaller, resulting in a smaller upper bound on the approximation bias.
However, as discussed in Section \ref{sec:appr}, the approximation bias can also be made small by choosing a large $q$ (roughly, $q_n \asymp O(\log(nK))$).
Therefore, $\beta_K$ and $q$ have similar roles.
We thus set $\beta_K=0.99$ to maintain a small variance and focus on the selection of $q$.

The selection of $\beta_1$ is essential for the variance of the PET method since a small $\beta_1$ may cause the variance of the PET estimator to be directly influenced by the variance of the standard IPW estimator ($\beta=0$).
One of the key motivations for the PET method is to variance reduction. To this end, we introduce an additional hyperparameter $\kappa$ ($0<\kappa\leq1$) and define a target variance as $\kappa\hat{\sigma}_{\text{IPW}}^2$.
Using $\kappa$, we propose the following Algorithm \ref{alg:select_beta_q} to select $\beta_1$ and $q$.
Note that when the augmentation estimator discussed in Section \ref{sec:aug} is used, Line 2 in Algorithm \ref{alg:select_beta_q} is replaced by the variance estimator of the standard AIPW estimator \citep{tsiatis2006semiparametric}, and the variance estimator in Line 6 is replaced by \eqref{avar3}.

\begin{algorithm}[ht!]
\caption{Selection of the hyperparameters $\beta_1$ and $q$ for implementing the PET approach.}
\label{alg:select_beta_q}
\begin{algorithmic}[1]
\State Set the candidates of $\beta_1$ and $q$ as $(\beta_{11},\dots,\beta_{1M_b})$ and $(q_1,\dots,q_{M_q})$.
Here, $\beta_{11}<\cdots<\beta_{1M_b}$ and $q_1<\cdots<q_{M_q}$ with $\beta_{1m_b}\in D$ and $q_{m_q}\in\mathbb{N}$.
\State Compute the variance estimator of the standard IPW estimator:\ $\hat{\sigma}_{\text{IPW}}^2$.
\State Set the target variance as $\kappa\hat{\sigma}_{\text{IPW}}^2$.
\For{each candidate value of $q_{m_q}$}
    \For{each candidate value of $\beta_{1m_b}$}
        \State Compute the variance estimator of the PET:\ $\hat{\sigma}^2_{m_q,m_b}$ using \eqref{avar} or \eqref{avar2}.
    \EndFor
    \If{$\min_{m_b\in\{1,\dots,M_b\}}\hat{\sigma}^2_{m_q,m_b}
    \geq \kappa\hat{\sigma}_{\text{IPW}}^2$}
        \State $m_b^*=M_b$
    \Else
        \State $m_b^*
        =
        \min\left\{
        m_b:\ \hat{\sigma}^2_{m_q,m_b}
        < \kappa\hat{\sigma}_{\text{IPW}}^2
        \right\}$
    \EndIf

    \State $\hat{\sigma}^2_{m_q}
    =
    \hat{\sigma}^2_{m_q,m_b^*}$
\EndFor
\If{$\min_{m_q\in\{1,\dots,M_q\}}\hat{\sigma}^2_{m_q}
    \geq \kappa\hat{\sigma}_{\text{IPW}}^2$}
        \State $m_q^*=1$
    \Else
        \State $m_q^*
        =
        \max\left\{
        m_q:\ \hat{\sigma}^2_{m_q}
        < \kappa\hat{\sigma}_{\text{IPW}}^2
        \right\}$
    \EndIf
    \State $\hat{\sigma}^2_{opt}
    =
    \hat{\sigma}^2_{m_q^*}$

\end{algorithmic}

\vspace{0.5em}
{\footnotesize{
Notes:\ $\beta_{1m_b}$ ($m_b=1,\dots,M_b$), $M_b$ candidate values of $\beta_1$; 
$q_{m_q}$ ($m_q=1,\dots,M_q$), $M_q$ candidate values of $q$. $\hat{\sigma}^2_{m_q,m_b}$, estimated variance of the PET method under $(\beta_{1},q)=(\beta_{1m_b},q_{m_q})$.
    }}
\end{algorithm}

The tuning parameter $\kappa$ admits a simple sample size interpretation.
Using Algorithm \ref{alg:select_beta_q}, it can be satisfied that
$$
\frac{\hat{\sigma}^2_{\text{opt}}}{n}\leq\kappa\frac{\hat{\sigma}^2_{\text{IPW}}}{n}=\frac{\hat{\sigma}^2_{\text{IPW}}}{n/\kappa}.
$$
In other words, the variance reduction achieved by Algorithm \ref{alg:select_beta_q} is at least equivalent to increasing the sample size of the IPW estimator from $n$ to $n/\kappa$.
For instance, $\kappa=5/6$ corresponds to a $20$\% increase in sample size.

%\subsection{Sensitivity analysis}\label{sec:sens}
\begin{rem}\label{rmk:SA}
Using Algorithm \ref{alg:select_beta_q}, we can select the hyperparameters $\beta_1$ and $q$ (with fixed $K$ and $\beta_K$ ($=0.99$)).
However, it is not necessarily clear whether the approximation bias in \eqref{assu:poly}, equivalently \eqref{apperr1}, is sufficiently small.
Therefore, it is recommended to conduct sensitivity analyses to examine how $\beta_1$ and $q$ impact the analysis results, based on the values selected by Algorithm \ref{alg:select_beta_q}. One option is to directly modify $(\beta_1,q)$ around the selected values.
This approach is simple and makes the sensitivity analysis results easy to interpret.
Another option is to plot the trajectories for several choices of $(\beta_1,q)$.
Specifically, using $\hat{\bld{\gamma}}$, we can plot the fitted model based on \eqref{mod1}:\ $\hat{\gamma}_0+\sum_{\ell=1}^{q}\hat{\gamma}_{\ell}\beta^{\ell}$. Together with the estimators for the beta weight family, these plots allow us to evaluate how much and how sensitively the PET method modifies the results. These sensitivity analyses are illustrated in the following real data applications.
\end{rem}

\begin{comment}
To assess this condition, we propose the following sensitivity analysis, particularly for the choice of $q$.
Specifically, for a fixed $q$, we consider decomposing a coefficient $\gamma_{k,q}$ as
$$
\gamma_{k,q}=\widetilde{\gamma}_{q}+\widetilde{\gamma}_{k,q}\ \ \ (k=1,\dots,K)
$$
and grouping the $K$ parameters into $\widetilde{K}$ strata ($\widetilde{K}<K$).
For instance, when $K=50$ and $\widetilde{K}=5$, every 10 values of $\widetilde{\gamma}_{k,q}$ are considered as approximately equal.
Under the decomposition, we re-estimate $\hat{\tau}_{P}$ with $\widetilde{\gamma}_{k,q}$ ($k=1,\dots,\widetilde{K}$).
By comparing the resulting estimates with the original estimator, we can assess the sensitivity of the PET estimator to violations of the approximation assumption underlying the choice of $q$.
\end{comment}

\section{Simulation experiments}\label{sec:simu}
We conduct simulation experiments under homogeneous and heterogeneous treatment effect settings, by modifying the settings in \cite{li2019addressing}. The number of simulation iterations is set to be 2,000 throughout. The key simulation settings and results are presented below, and additional simulation materials are provided in Appendix \ref{app:add_sim}.

\subsection{Simulation design}
We consider data generating mechanisms similar to the settings in \cite{li2019addressing}.
The marginal treatment prevalence is approximately 0.2 (${\rm E}[A]\approx 0.2$), and the ATE is 0.75 under both the homogeneous and heterogeneous treatment effect settings.
Full details of the data generating mechanisms are provided in Appendix \ref{app:add_sim}.

We mainly compare six methods:\ the standard IPW estimator, the IPW estimator for the ATO using overlap weights (OW) ($w_{1}(x)=e(x)(1-e(x))$), Crump's trimmed weighting method with the optimal cutoff value (using \texttt{PStrim} function in the \texttt{PSweight} package), the smooth weighting method \citep{yang2018asymptotic}, the bias-corrected method \citep{ma2020robust}, and the PET method.
For all methods, the propensity score is estimated using the correctly specified model.
For the bias-corrected method, we adapt their publicly available replication code, originally implemented for the ATT, to estimate the ATE.
For the PET methods, we set $K=50$ and $\beta_K=0.99$ in all settings.
The candidate values of $\beta_1$ are $\{0.44, 0.49, \dots, 0.69\}$ and $q$ are $\{1,2,3,4\}$.
Furthermore, the AIPW versions of these methods, except for the bias-corrected method are also considered. We evaluate three situations:\ both models are correctly specified, only the propensity score model is correctly specified, or only the outcome model is correctly specified.
For the misspecified model, we exclude the covariate $X_1$ from the true propensity score or outcome model.

We evaluated the methods in terms of absolute bias, empirical standard error (EmpSE), ratio of the estimated standard error to the empirical standard error (EstSE/EmpSE), coverage probability (CP), and boxplots of the estimated ATE based on 2,000 simulation iterations.
The EstSE is calculated using the corresponding asymptotic variance.
The CP is defined as the proportion of 95\% confidence intervals (CI) that contain the true $\tau$.

\subsection{Simulation results}
The results are summarized in Tables \ref{tab1} and \ref{tab:app4}.
In the main manuscript, we only consider the results under the limited overlap scenario with the heterogeneous treatment effect setting. The standard IPW results exhibit large biases and EmpSEs, resulting in large RMSEs in both small sample ($n=500$) and large sample ($n=2000$) settings, indicating unstable performance. In contrast, the results for the ATO, Crump's method, and smooth weighting method show stable performance in terms of EmpSE. However, there are nonignorable biases to the true ATE even in the large sample setting, also resulting in large RMSEs.
The bias-corrected method has larger bias than the standard IPW estimator, but smaller EmpSE, resulting in a smaller RMSE.
One possible reason for the large bias is the sparse support in the propensity score tail regions used to fit the local extrapolation models.
For the proposed PET methods, the bias is smaller than those of the ATO, Crump's method, and Yang's method, while the EmpSE is smaller than that of standard IPW estimator, at least when $q=1$, $2$, or optimally selected. Regarding the selection of $\beta_1$ and $q$, Algorithm \ref{alg:select_beta_q} can identify a reasonable combination; increasing the accuracy compared with the standard IPW results.
Additionally, among the candidate values of $q$, smaller values tend to yield smaller EmpSEs but larger biases than larger values in the large sample setting.
This is consistent with the discussion in Sections \ref{sec:appr} and \ref{sec:hyp}. Finally, compared with the undercoverage of the standard IPW estimator, that of PET is improved. This can be explained by the poor variance estimation of the standard IPW estimator, as reflected by the EstSE/EmpSE ratio. The EstSE/EmpSE ratio for the PET estimator with hyperparameter selection is 89.9\%, compared with 58.4\% for the standard IPW estimator in the large sample setting.
Therefore, the variance estimator for the standard IPW estimator is also unreliable under this setting, and the variance estimator under PET appears more credible.

Turning to the AIPW estimators, the proposed AIPW-PET method shows a tendency similar to that of the IPW-based estimator.
Additionally, the difference between AIPW-PET and the standard AIPW estimator becomes smaller in the large sample setting than in the small sample setting; this tendency is especially observed in the strong overlap scenario (Appendix \ref{app:add_sim}).
Even when either the propensity score or outcome model is misspecified, both the standard AIPW and AIPW-PET estimators show unbiased results, especially in the large sample setting.
In this sense, both estimators exhibit double robustness in our simulation settings. A notable point, however, is that AIPW-PET has smaller EmpSEs, resulting in smaller RMSEs than the standard AIPW estimator. As mentioned in Remark \ref{rmk2} in Section \ref{sec:aug}, this represents a performance difference between the two estimators, and AIPW-PET can potentially reduce the variance of the estimator for the ATE.
Overall, the PET methods can estimate the ATE accurately even in limited overlap settings where the standard estimators for the ATE perform poorly.

\begin{table}[ht!]
\begin{center}
\caption{Results of the simulation experiment under the limited overlap scenario with the heterogeneous treatment effect. The number of iterations is 2,000.}\label{tab1}
    \scalebox{0.9}{$
    \begin{tabular}{cc|ccccc|ccccc}
    \midrule
       \multicolumn{10}{l}{\textbf{IPW estimators}} \\
       \addlinespace[0.3em]
       \hline
Method & $q$ & \multicolumn{5}{c|}{$n=500$} & \multicolumn{5}{c}{$n=2000$} \\ 
        & & $|$bias$|$ & EmpSE & RMSE & $\frac{\mathrm{EstSE}}{\mathrm{EmpSE}}$ & CP 
        & $|$bias$|$ & EmpSE & RMSE & $\frac{\mathrm{EstSE}}{\mathrm{EmpSE}}$ & CP \\ \hline
IPW
& --
& 0.455 & 1.083 & 1.175 & 0.463 & 50.9
& 0.190 & 0.789 & 0.812 & 0.581 & 65.3 \\

OW
& --
& 0.361 & 0.241 & 0.434 & 0.963 & 64.9
& 0.367 & 0.113 & 0.384 & 1.027 & 10.5 \\

Trimmed
& --
& 0.408 & 0.290 & 0.501 & 0.903 & 63.2
& 0.416 & 0.135 & 0.437 & 0.992 & 12.8 \\

Smooth
& --
& 0.390 & 0.311 & 0.499 & 1.144 & 83.0
& 0.407 & 0.148 & 0.433 & 1.069 & 24.7 \\

Bias-corrected
& --
& 0.545 & 0.687 & 0.877 & 0.848 & 90.9
& 0.351 & 0.498 & 0.609 & 1.010 & 85.7 \\

PET
& selected
& 0.104 & 0.566 & 0.576 & 0.795 & 93.8
& 0.088 & 0.427 & 0.436 & 0.898 & 92.2 \\

& 1
& 0.119 & 0.557 & 0.570 & 0.804 & 94.4
& 0.112 & 0.346 & 0.364 & 0.854 & 93.8 \\

& 2
& 0.008 & 1.033 & 1.033 & 0.588 & 78.6
& 0.034 & 0.522 & 0.523 & 0.828 & 86.4 \\

& 3
& 0.120 & 1.500 & 1.505 & 0.457 & 64.6
& 0.033 & 0.766 & 0.767 & 0.707 & 78.0 \\

& 4
& 0.263 & 1.632 & 1.653 & 0.376 & 53.5
& 0.080 & 0.960 & 0.963 & 0.572 & 69.1 \\
       \hline
       \midrule
       \multicolumn{10}{l}{\textbf{AIPW estimators}} \\
       \addlinespace[0.3em]
       \hline
Method & $q$ & \multicolumn{5}{c|}{$n=500$} & \multicolumn{5}{c}{$n=2000$} \\ 
        & & $|$bias$|$ & EmpSE & RMSE & $\frac{\mathrm{EstSE}}{\mathrm{EmpSE}}$ & CP 
        & $|$bias$|$ & EmpSE & RMSE & $\frac{\mathrm{EstSE}}{\mathrm{EmpSE}}$ & CP \\ \hline
          AIPW & -- & 0.008 & 0.480 & 0.480 & 0.647 & 80.0 & 0.007 & 0.303 & 0.303 & 0.735 & 88.4 \\
          Aug.\ OW & -- & 0.360 & 0.238 & 0.432 & 0.945 & 62.7 & 0.366 & 0.112 & 0.383 & 1.007 & 9.9 \\
      Aug.\ trimmed & -- & 0.396 & 0.270 & 0.479 & 0.935 & 63.9 & 0.408 & 0.129 & 0.428 & 0.997 & 11.3 \\
    Aug.\ smooth & -- & 0.396 & 0.263 & 0.475 & 0.946 & 62.5 & 0.407 & 0.126 & 0.426 & 1.000 & 10.8 \\
          AIPW-PET & selected & 0.062 & 0.371 & 0.376 & 0.780 & 83.6 & 0.023 & 0.236 & 0.237 & 0.890 & 91.0 \\
              & 1 & 0.138 & 0.330 & 0.358 & 0.896 & 87.7 & 0.117 & 0.174 & 0.210 & 0.953 & 86.4 \\
              & 2 & 0.051 & 0.431 & 0.434 & 0.813 & 88.1 & 0.027 & 0.233 & 0.235 & 0.913 & 93.2 \\
              & 3 & 0.011 & 0.542 & 0.542 & 0.731 & 85.5 & 0.009 & 0.283 & 0.283 & 0.851 & 93.0 \\
              & 4 & 0.018 & 0.583 & 0.583 & 0.680 & 83.8 & 0.003 & 0.312 & 0.312 & 0.801 & 92.6 \\
       \hline
    \end{tabular}
       $}
\end{center}
    {\footnotesize{
    Notes:\ EmpSE, empirical standard error; RMSE, root mean squared error; EstSE, mean estimated standard error; CP, coverage probability of the 95\% confidence interval.
    For Crump’s trimming method, \(\alpha\) is selected using the optimal cutoff implemented in the \texttt{PStrim} function in the \texttt{PSweight} package.
    For Yang’s smooth weighting method, \(\alpha\) is set equal to the cutoff selected for Crump’s method, with $\epsilon=10^{-2}$.
    For PETs, $K=50$, $\beta_K=0.99$, and $\kappa=5/6$ (for the PET) or $\kappa=1$ (for the AIPW-PET). Candidate values of $\beta_1$ are $\{0.44, 0.49, \dots, 0.69\}$ and $q$ are $\{1,2,3,4\}$.
    For the bias-corrected method, a confidence interval based on studentized subsampling is used.
    For PET and AIPW-PET, confidence intervals use the variance estimators in \eqref{avar2} and \eqref{avar3}, respectively. For the remaining IPW estimators, robust variance estimators accounting for propensity score estimation are used.
    }}
\end{table}

\begin{table}[ht!]
\begin{center}
\caption{Results of the simulation experiment with model misspecification. The number of iterations is 2,000.}\label{tab:app4}
    \begin{tabular}{c|ccccc|ccccc}
    \midrule
       \multicolumn{10}{l}{\textbf{Outcome model misspecification}} \\
       \addlinespace[0.3em]
       \hline
       Method & \multicolumn{5}{c|}{$n=500$} & \multicolumn{5}{c}{$n=2000$} \\ 
        & $|$bias$|$ & EmpSE & RMSE & $\frac{\mathrm{EstSE}}{\mathrm{EmpSE}}$ & CP 
        & $|$bias$|$ & EmpSE & RMSE & $\frac{\mathrm{EstSE}}{\mathrm{EmpSE}}$ & CP \\ \hline
       AIPW & 0.079 & 0.534 & 0.540 & 0.621 & 80.3 & 0.047 & 0.362 & 0.365 & 0.676 & 86.4 \\
       AIPW-PET & 0.040 & 0.375 & 0.377 & 0.835 & 87.6 & 0.018 & 0.261 & 0.262 & 0.885 & 92.9 \\
       \hline
       \midrule
       \multicolumn{8}{l}{\textbf{Propensity score model misspecification}} \\
       \addlinespace[0.3em]
       \hline
       Method & \multicolumn{5}{c|}{$n=500$} & \multicolumn{5}{c}{$n=2000$} \\ 
        & $|$bias$|$ & EmpSE & RMSE & $\frac{\mathrm{EstSE}}{\mathrm{EmpSE}}$ & CP 
        & $|$bias$|$ & EmpSE & RMSE & $\frac{\mathrm{EstSE}}{\mathrm{EmpSE}}$ & CP \\ \hline
       AIPW & 0.008 & 0.487 & 0.487 & 0.634 & 79.8 & 0.002 & 0.300 & 0.300 & 0.728 & 88.2 \\
       AIPW-PET & 0.062 & 0.376 & 0.381 & 0.767 & 82.9 & 0.025 & 0.237 & 0.238 & 0.871 & 90.1 \\
       
       \hline
    \end{tabular}
\end{center}
    {\footnotesize{
    Notes:\ EmpSE, empirical standard error; RMSE, root mean squared error; EstSE, mean estimated standard error; CP, coverage probability of the 95\% confidence interval.
    For AIPW-PET, $K=50$, $\beta_K=0.99$, and $\kappa=1$. Candidate values of $\beta_1$ are $\{0.44, 0.49, \dots, 0.69\}$ and $q$ are $\{1,2,3,4\}$.
    For AIPW-PET, confidence intervals use the variance estimators in \eqref{avar3}.
    }}
\end{table}

\section{Two real data applications} \label{sec:appl}

\subsection{Impact of right heart catheterization on mortality}
\subsubsection{Background}\label{sec7.1}
It has been reported that Right Heart Catheterization (RHC) has a harmful effect even after adjusting for sufficient confounders \citep{connors1996effectiveness, chen2020right}.
Here, we conduct a reanalysis using the dataset available from the website:\ \url{https://hbiostat.org/data/}.
Additional materials are provided in Appendix \ref{app:realdata}. In this dataset, there are 5,735 individuals. In the following data analysis, the assignment variable is defined as whether RHC was performed, and the outcome variable is death within 30 days after admission.
The number and proportion of subjects receiving RHC are 2,184 (38.1\%), and the numbers and proportions of deaths are 830 (38.0\%) in the treatment group and 1,088 (30.6\%) in the control group, respectively.
The difference in sample means between the groups is 0.074 (95\% CI:\ 0.049, 0.099).
In this analysis, we use 48 potential confounders without missing values among the original 53 confounders, since the remaining five variables have substantial missingness and limited information \citep{harada2025false}. Based on the 48 confounders, the propensity score is estimated using a standard logistic regression model with linear terms. Compared with the simulation experiment conducted in Section \ref{sec:simu}, the estimated propensity scores indicate moderate overlap.

\subsubsection{Results}
% The results are summarized in Table \ref{tab3} and demonstrate the same scientific conclusion:\ after confounder adjustment, RHC leads to increased chance of mortality.
% Similar to the results of the simulation experiments, the standard IPW estimator has the widest CI, whereas the overlap weighting estimator for the ATO has the narrowest one.
% For the proposed PET method, using Algorithm \ref{alg:select_beta_q}, $q=1$ and $\beta_1=0.69$ (the most conservative choices) are selected.
% The CI length of the PET method is similar to that of Crump's trimming method, while its point estimate is similar to that of the standard IPW estimator.
% Therefore, as expected, the PET method can estimate the ATE with improved precision.

The results are summarized in Table \ref{tab3}. Under the causal identification assumptions, all adjusted analyses suggest that RHC increases the risk of 30-day mortality by approximately 5-6 percentage points, with all 95\% confidence intervals excluding zero. The standard IPW estimator has a relatively wide confidence interval, whereas  the ATO (OW) has the narrowest. The latter comparison should be interpreted cautiously, however, because overlap weighting and trimming improve precision partly by targeting populations different from that of the canonical ATE. 
The bias-corrected method shows a similar tendency; however, its confidence interval is the widest among the competing methods.
For the proposed PET method, Algorithm \ref{alg:select_beta_q} selects $q=1$ and $\beta_1=0.69$. The resulting estimate is 0.055 (95\% CI:\ 0.027, 0.083), which is close to the standard IPW estimate of 0.053 but has a confidence interval approximately 7\% shorter. Its precision is also comparable to that of Crump's trimming method, while PET continues to target the ATE under the polynomial approximation condition. Estimates obtained using $q=2,3,$ and $4$ are nearly identical, suggesting that the substantive conclusion is not sensitive to the polynomial degree. Overall, PET reduces some of the variability associated with standard IPW without materially changing the estimated treatment effect. The precision gain is modest, as expected given the moderate degree of propensity score overlap in this application.

\begin{table}[ht!]
\begin{center}
\caption{Results of the IPW estimators in the RHC data.}\label{tab3}
    \begin{tabular}{cc|cc}
       \hline
       Method & $q$ & Point estimate (95\%CI) & CI length \\ \hline
       % Crude & -- & 0.074 (0.049, 0.099) & 0.050 \\
       IPW & -- & 0.053 (0.024, 0.083) & 0.059 \\
       OW & -- & 0.059 (0.033, 0.085) & 0.052 \\
       Trimmed & -- & 0.056 (0.028, 0.084) & 0.055 \\
       Smooth & -- & 0.058 (0.029, 0.087) & 0.057 \\
       Bias-corrected & -- & 0.046 (0.011, 0.083) & 0.072 \\
       PET & $1$ (selected) & 0.055 (0.027, 0.083) & 0.055 \\
           & $2$ & 0.054 (0.025, 0.083) & 0.058 \\
           & $3$ & 0.054 (0.024, 0.083) & 0.058 \\
           & $4$ & 0.053 (0.025, 0.082) & 0.057 \\
       \hline
    \end{tabular}
\end{center}
    {\footnotesize{
    Notes:\ For all methods, we use the implementation settings specified in the simulation study.
    }}
\end{table}

%\subsubsection{Sensitivity analyses}
% We conduct the sensitivity analyses following Remark \ref{rmk:SA}. First, we confirm the sensitivity to the choice of $(\beta_1,q)$.
% When $q=1$, the results for $\beta_1=0.29$ and $\beta_1=0.14$ are 0.054 (0.026, 0.083) and 0.054 (0.026, 0.083), respectively.
% When $q=2$, the results for $\beta_1=0.29$ and $\beta_1=0.14$ are 0.054 (0.025, 0.083) and 0.054 (0.025, 0.083), respectively.
% For $q=3$ and $4$, the same tendency is observed.
% Therefore, the results are expected to be robust to the choice of $(\beta_1,q)$ in this application. Compared with the IPW estimators for the beta weight family, the polynomial regressions in our proposed PET procedure can capture their point estimates well (Figure \ref{zu_traj}).
% Additionally, their trajectories are almost consistent across different values of $(\beta_1,q)$.

\begin{figure}[ht!]
\begin{center}
\begin{tabular}{c}
\includegraphics[width=17cm]{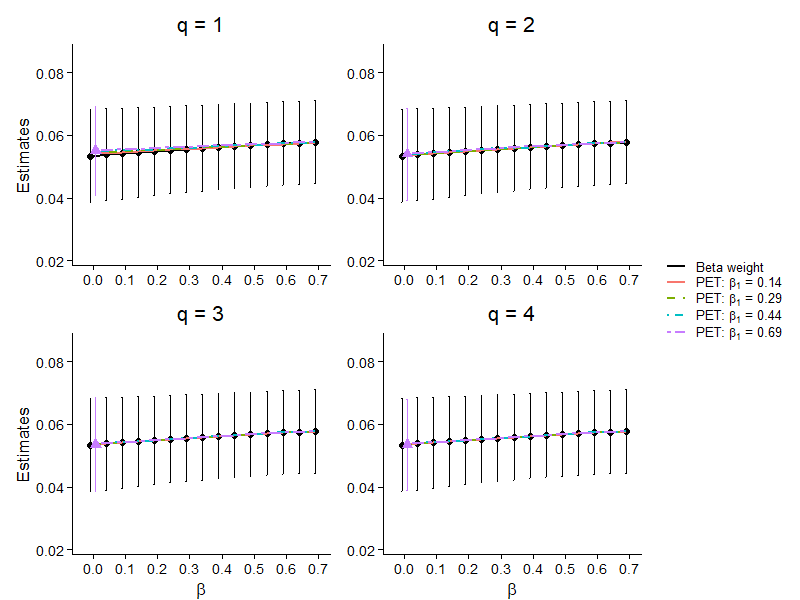}
\end{tabular}\caption{Trajectory of the IPW estimators for the beta weight family and polynomial regressions in the RHC data. The error bars represent the estimated standard errors. For all PET methods, $\beta_1=0.69$ is selected. The IPW estimator for the beta weight family at $\beta=0$ (the leftmost one) is consistent with the standard IPW estimator.}
\label{zu_traj}
\end{center}
\end{figure}

We further conduct the sensitivity analyses described in Remark \ref{rmk:SA} to assess the influence of $(\beta_1,q)$. When $q=1$, the results for $\beta_1=0.29$ and $\beta_1=0.14$ are both 0.054 (95\% CI:\ 0.026, 0.083). When $q=2$, the corresponding results are both 0.054 (95\% CI:\ 0.025, 0.083), with similarly stable results for $q=3$ and $4$. Figure \ref{zu_traj} further shows that the fitted polynomial curves closely track the constituent IPW estimates across the beta weight family and yield nearly identical extrapolated values across the considered specifications. These findings support the adequacy of a low-order polynomial approximation in this application and indicate that the PET estimate is insensitive to the choices of $\beta_1$ and $q$. This insensitivity is possibly because there exists limited treatment effect heterogeneity due to RHC.

\subsection{Impact of smoking habits on forced expiratory volume}
\subsubsection{Background}
The effect of smoking during childhood and adolescence on pulmonary function, as measured by forced expiratory volume (FEV), has been considered as an example of causal inference under limited overlap \citep{matsouaka2024causal}. We reanalyze the \texttt{fev} dataset from the \texttt{doBy} package; additional details are provided in Appendix \ref{app:realdata}. Because none of the participants younger than 9 years reported smoking, we restrict the analysis to 439 participants aged 9--19 years, thereby excluding an age range in which the positivity assumption is structurally violated. The exposure is an indicator of smoking status, and the outcome is FEV, treated as a continuous variable. Among the 439 participants, 65 (14.8\%) reported smoking. The mean (SD) FEV was 3.28 (0.75) among smokers and 2.97 (0.76) among nonsmokers, corresponding to an unadjusted mean difference of 0.304 (95\% CI:\ 0.102, 0.505). This crude difference should not be interpreted causally because smoking status is associated with important determinants of FEV, particularly age and height. We adjust for the three measured potential confounders available in the dataset (age, sex, and height) and estimate the propensity score using a logistic regression model containing their linear terms. Because only these three covariates are available, residual confounding remains a potential concern and the results should be interpreted as illustrative \citep{matsouaka2024causal}. The estimated propensity score distributions exhibit limited overlap between smokers and nonsmokers. Thus, in contrast to the moderate overlap observed in the RHC application, this dataset provides a more challenging and complementary setting in which to evaluate the performance of PET.

\subsubsection{Results}
The results are summarized in Table \ref{tab3_fev}. In contrast to the positive unadjusted mean difference of 0.304, all adjusted point estimates are negative, ranging from $-0.041$ to $-0.568$. This reversal underscores the importance of adjusting for differences between smokers and nonsmokers in age, height, and sex. Nevertheless, except for the bias-corrected method, all 95\% confidence intervals include zero and remain compatible with a range of effects. Thus, although the adjusted estimates consistently suggest that smoking may reduce FEV, the data do not support a definitive conclusion regarding either the presence or the magnitude of this effect. The variation across methods should be interpreted in light of their different target populations. The ATO (OW), trimmed weighting, and smooth weighting estimators place greater emphasis on participants with better covariate overlap and yield estimates closer to zero with substantially narrower confidence intervals. In contrast, the standard IPW estimator targets the ATE in the full population and produces a substantially wider confidence interval than these overlap-focused estimators, reflecting the difficulty of estimating this effect under limited overlap and with only 65 smokers. Differences among these point estimates may be compatible with treatment effect heterogeneity across covariate regions, but they may also reflect sparse support and sampling variability.

%\subsubsection{Sensitivity analyses}

For the PET method, Algorithm \ref{alg:select_beta_q} selects $(\beta_1,q)=(0.44,1)$. The resulting estimate is $-0.080$ (95\% CI:\ $-0.373$, 0.213), with a confidence interval approximately 19\% shorter than that of the standard IPW estimator. However, the PET point estimate is less negative than the standard IPW estimate of $-0.184$, illustrating that the gain in stability is accompanied by greater reliance on polynomial extrapolation. When $q\geq2$, the PET estimates generally move toward the standard IPW estimate, but their confidence intervals are similar to or wider than that of standard IPW. This pattern is consistent with the small sample results in Table \ref{tab1}:\ a low-degree polynomial can stabilize the extrapolation, whereas additional polynomial terms provide greater flexibility at the cost of increased estimation variability.

\begin{table}[ht!]
\begin{center}
\caption{Results of the IPW estimators in the FEV data.}\label{tab3_fev}
    \begin{tabular}{cc|cc}
       \hline
       Method & $q$ & Point estimate (95\%CI) & CI length \\ \hline
       % Crude & -- & 0.304 (0.102, 0.505) & 0.403 \\
       IPW & -- & $-$0.184 ($-$0.545, 0.177) & 0.722  \\
       OW & -- & $-$0.122 ($-$0.282, 0.037) & 0.319 \\
       Trimmed & -- & $-$0.041 ($-$0.209, 0.128) & 0.337 \\
       Smooth & -- & $-$0.060 ($-$0.277, 0.157) & 0.434 \\
       Bias-corrected & -- & $-$0.568 ($-$1.192, $-$0.212) & 0.980 \\
       PET & $1$ (selected) & $-$0.080 ($-$0.373, 0.213) & 0.586 \\
           & $2$ & $-$0.126 ($-$0.491, 0.238) & 0.730 \\
           & $3$ & $-$0.183 ($-$0.583, 0.217) & 0.800 \\
           & $4$ & $-$0.198 ($-$0.560, 0.164) & 0.725 \\
       \hline
    \end{tabular}
\end{center}
    {\footnotesize{
    Notes:\ For all methods, we use the implementation settings specified in the simulation study.
    }}
\end{table}

\begin{figure}[ht!]
\begin{center}
\begin{tabular}{c}
\includegraphics[width=17cm]{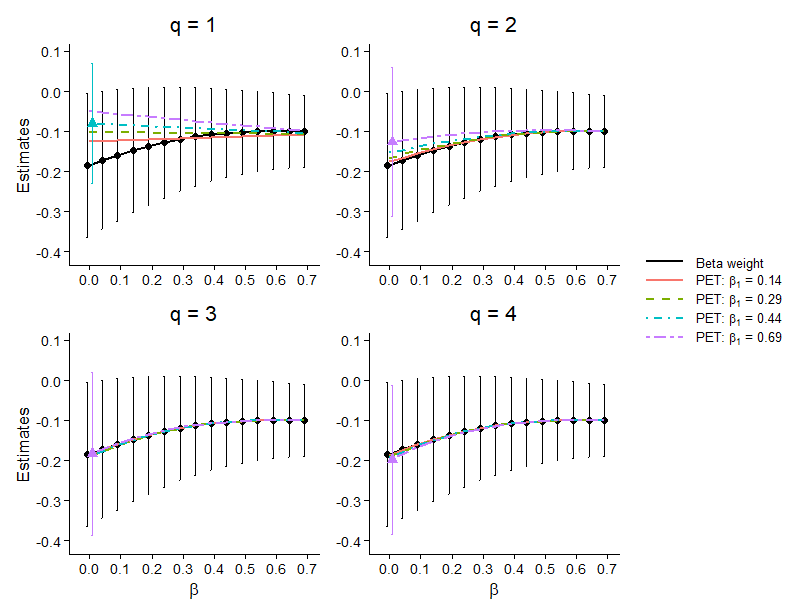}
\end{tabular}\caption{Trajectory of the IPW estimators for the beta weight family and polynomial regressions in the FEV data. The error bars represent the estimated standard errors. For the PET methods, $\beta_1=0.44$ is selected when $q=1$. For the other $q$, $\beta_1=0.69$ is selected. The IPW estimator for the beta weight family at $\beta=0$ (the leftmost one) is consistent with the standard IPW estimator.}
\label{zu_traj_FEV}
\end{center}
\end{figure}

We next examine the sensitivity of the PET results to $(\beta_1,q)$. When $q=1$, the estimates for $\beta_1=0.29$ and $\beta_1=0.14$ are $-0.102$ (95\% CI:\ $-0.412$, 0.208) and $-0.125$ (95\% CI:\ $-0.450$, 0.200), respectively. Their corresponding confidence interval lengths are 0.621 and 0.650, both of which remain shorter than the confidence interval from standard IPW. Thus, as $\beta_1$ decreases, the PET estimate moves toward the standard IPW estimate, while its confidence interval becomes wider. This pattern reflects a central trade-off in PET. That is, choosing a smaller $\beta_1$ reduces the distance over which the WATE trajectory must be extrapolated to $\beta=0$, but it also incorporates WATE estimators that are more sensitive to limited overlap. When $q=2$, the estimates for $\beta_1=0.29$ and $\beta_1=0.14$ are $-0.166$ (95\% CI:\ $-0.538$, 0.206) and $-0.175$ (95\% CI:\ $-0.542$, 0.191), respectively, with confidence interval lengths of 0.744 and 0.733. Similar patterns are observed for $q=3$ and $4$. Overall, smaller values of $\beta_1$ and larger values of $q$ produce estimates closer to the standard IPW estimate, but they provide little or no precision gain. Although the direction of the estimated effect remains negative across the considered specifications, its magnitude is moderately sensitive to the hyperparameter choices. The trajectory plots further clarify these findings (Figure \ref{zu_traj_FEV}). In contrast to the nearly linear and stable trajectory observed in the RHC application, the WATE trajectory in the FEV data exhibits appreciable curvature. A first-order polynomial does not follow this curvature as closely as higher-order polynomials, but it yields a more stable extrapolation. Conversely, increasing $q$ improves the fit to the WATE estimates over the observed range of $\beta$ but does not necessarily improve estimation at $\beta=0$, which lies outside that range. Indeed, the additional flexibility can amplify uncertainty in the extrapolated value. The trajectory plots should therefore be viewed as diagnostic tools for assessing sensitivity to $(\beta_1,q)$ rather than as a means of identifying a ``correct'' polynomial model.

Taken together, the two applications illustrate complementary operating regimes of PET. In the RHC data, where overlap is moderate and the WATE trajectory is nearly linear, the PET results are largely insensitive to the hyperparameter choices. In the FEV data, where overlap is limited and the exposed group is small, the extrapolated estimate is more sensitive to both the starting value $\beta_1$ and the polynomial degree $q$. These findings motivate a two-step workflow for applied use of PET. First, Algorithm \ref{alg:select_beta_q} can be used to obtain an initial data-driven choice of $(\beta_1,q)$. Second, this primary analysis should be accompanied by sensitivity analyses over a reasonable range of hyperparameter values, together with trajectory plots showing how the WATE estimates are translated into the extrapolated ATE estimate.

\section{Discussion}
Estimating the canonical ATE under limited overlap is intrinsically difficult because inverse probability weighting assigns the greatest leverage to observations in sparsely represented covariate regions, producing high variability and potentially unreliable inference. Existing weighting and trimming strategies improve stability by targeting populations with stronger empirical support, but thereby change the scientific question. PET takes a different route. We show that the WATEs in the beta weight family form a structured estimand path anchored at $\tau_0=\tau$ and admit a polynomial expansion in $\beta$. PET estimates the more stable WATEs at positive values of $\beta$ and extrapolates this path to $\beta=0$. Thus, WATEs serve as building blocks for estimating the canonical ATE rather than as alternative final targets. This connection across causal estimands is the central novelty of PET and provides a new way to retain the ATE as the target while exploiting the greater stability of WATE estimation.

The proposed approach relies on a polynomial approximation condition, whose exact form can be stringent. This condition makes explicit the unavoidable extrapolation required when parts of the target population are weakly supported by the observed data. Our Taylor expansion isolates the approximation remainder, and the asymptotic theory requires only that the resulting approximation bias be negligible relative to sampling uncertainty; Section \ref{sec:appr} provides theoretical and practical arguments for why this requirement can be plausible. The simulations offer strong empirical support for these arguments:\ under limited overlap, PET substantially reduces variability relative to standard IPW while retaining considerably less bias than methods that directly target alternative populations, and the proposed trajectory plots and sensitivity analyses reveal when conclusions depend on the approximation. PET is therefore particularly useful when the canonical ATE is dictated by the scientific or policy question and redefining the target population is not satisfactory. More broadly, PET offers a complementary perspective in which the ATE and WATEs are connected points along an estimand path, allowing more stable estimands to inform inference about a challenging but substantively preferred target.

\noindent

\vspace{0.4cm}
{\bf Acknowledgement:}\ This work was supported by JSPS KAKENHI Grant Numbers 24K20742 and 25K21166.\\
\noindent
{\bf Generative AI use:}\ 
ChatGPT (OpenAI) was used primarily for English language refinement and software coding assistance. All AI-assisted text and code were reviewed and verified by the authors.\\
\noindent
{\bf Supporting information:}\ An R function to reproduce the simulation experiments is available at \url{https://github.com/SOrihara/PET_method}.

%   Reference
\bibliography{ref}

%%%%%%%%%%%%%%%%%%%%%%%%%%%%%%%%%%%%%%%%%%%%%%%%%%%%%%%%%%%%%%%%%%%%%%%%%%%%%%%%%%

\appendix
\setcounter{table}{0}
\renewcommand{\thetable}{A.\arabic{table}}

\setcounter{figure}{0}
\renewcommand{\thefigure}{A.\arabic{figure}}

\newpage
\section{Proofs}\label{appa}
\subsection{Proof of Proposition \ref{prop2}}
Assume that
\begin{align}
{\rm E}\left[|\log \{e(X)(1-e(X))\}|^{q+1}\right]<\infty. \label{assu_logm}
\end{align}

Note that the moment condition \eqref{assu_logm} is weaker than the strict overlap assumption, in the sense that strict overlap implies this condition, whereas the converse does not necessarily hold.

Define $\alpha(x)=e(x)(1-e(x))$, and note that under the positivity assumption, $|\log\alpha(x)|<\infty$ for each $x\in\mathcal{X}$.
To prove the proposition, it is sufficient to show that for all $r=0,1,\dots,q+1$, there exists $C_r<\infty$ such that, for all $x\in\mathcal{X}$,
\begin{align}
\sup_{\eta\in[0,\beta]}\left|h_{\eta}^{(r)}(x)\right|\leq C_r(1+|\log \alpha(x)|^r). \label{app_logm}
\end{align}

When $r=0$, since $\eta\leq\beta$ and $\alpha(x)\leq 1/4$,
$$
h_{\eta}^{(0)}(x)=\frac{\alpha(x)^{\eta}}{{\rm E}[\alpha(X)^{\eta}]}\leq\frac{1}{{\rm E}[\alpha(X)^{\beta}]}=:C_0,
$$
where $C_0<\infty$.

Suppose that \eqref{app_logm} holds for $r=0,1,\dots,r'-1$. 
When $r=r'$, from the generalized Leibniz rule,
\begin{align*}
\alpha(x)^{\eta}\{\log\alpha(x)\}^{r'}&=\sum_{\ell=0}^{r'}\binom{r'}{\ell}h^{(r'-\ell)}_{\eta}(x){\rm E}\left[\alpha(X)^{\eta}\{\log\alpha(X)\}^{\ell}\right].
\end{align*}
Here, note that for $\ell=0,1,\dots,q+1$ and $\eta\in[0,\beta]$,
$$
{\rm E}\left[|\alpha(X)^{\eta}\{\log\alpha(X)\}^{\ell}|\right]\leq{\rm E}\left[|\log\alpha(X)|^{\ell}\right]\leq{\rm E}\left[|\log\alpha(X)|^{q+1}\right]<\infty.
$$
Then,
$$
h^{(r')}_{\eta}(x)=\frac{1}{{\rm E}[\alpha(X)^{\eta}]}\left[\alpha(x)^{\eta}\{\log\alpha(x)\}^{r'}-\sum_{\ell=1}^{r'}{}\binom{r'}{\ell}h^{(r'-\ell)}_{\eta}(x){\rm E}\left[\alpha(X)^{\eta}\{\log\alpha(X)\}^{\ell}\right]\right].
$$
Therefore,
\begin{align*}
\left|h^{(r')}_{\eta}(x)\right|
&\leq
C_0\left[
\left|\log\alpha(x)\right|^{r'}+\sum_{\ell=1}^{r'}{}\binom{r'}{\ell}\left\{
C_{r'-\ell}(1+|\log \alpha(x)|^{r'-\ell})
\right\}{\rm E}\left[|\log\alpha(X)|^{\ell}\right]
\right] \\
&\leq
C_0\left|\log\alpha(x)\right|^{r'}+C_0\max_{\ell=1,\dots,r'}\left\{\binom{r'}{\ell}
C_{r'-\ell}{\rm E}\left[|\log\alpha(X)|^{\ell}\right]\right\}\sum_{\ell=1}^{r'}{}(1+|\log \alpha(x)|^{r'-\ell}) \\
&=
C_0\left|\log\alpha(x)\right|^{r'}+C'_{r'}\sum_{\ell=1}^{r'}{}(1+|\log \alpha(x)|^{r'-\ell}) \\
&\leq
C^{''}_{r'}\left\{\left|\log\alpha(x)\right|^{r'}+\sum_{\ell=1}^{r'}{}(1+|\log \alpha(x)|^{r'-\ell})\right\} \\
&\leq
C_{r'}(1+|\log \alpha(x)|^{r'}),
\end{align*}
where $C_{r'}<\infty$.
By induction, \eqref{app_logm} holds for all $r=0,1,\dots,q+1$.

Actually, by \eqref{assu_logm} and \eqref{app_logm},
\begin{align*}
{\rm E}\left[\left|\left.h_{\beta}^{(q+1)}(X)\right|_{\beta=c(X)}\tau(X)\right|\right]
&\leq
\sup_{x\in\mathcal{X}}|\tau(x)|{\rm E}\left[\left|\left.h_{\beta}^{(q+1)}(X)\right|_{\beta=c(X)}\right|\right] \\
&\leq
C_{q+1}\sup_{x\in\mathcal{X}}|\tau(x)|{\rm E}\left[(1+|\log \alpha(X)|^{q+1})\right] \\
&<\infty.
\end{align*}

\subsection{Proof of Theorem \ref{theo1}}
For $\hat{\tau}_P$,
\begin{align*}
\hat{\tau}_P&=\frac{\bld{1}_K^{\top}\bar{P}_B\hat{\bld{\tau}}}{\bld{1}_K^{\top}\bar{P}_B\bld{1}_K}=\frac{1}{\bld{1}_K^{\top}\bar{P}_B\bld{1}_K}\left\{\sum_{k=1}^{K}\hat{\tau}_{\beta_k}-\bld{1}_K^{\top}\left(
\begin{array}{ccc}
b_{11}&\cdots&b_{1K}\\
\vdots&\ddots&\vdots\\
b_{K1}&\cdots&b_{KK}
\end{array}
\right)\left(
\begin{array}{c}
\hat{\tau}_{\beta_1}\\
\vdots\\
\hat{\tau}_{\beta_K}
\end{array}
\right)\right\}\\
&=\frac{1}{\bld{1}_K^{\top} \bar{P}_B\bld{1}_K}\sum_{k=1}^{K}\left\{1-\sum_{k'=1}^{K}b_{k'k}\right\}\hat{\tau}_{\beta_k}\\
&=\frac{1}{n}\sum_{i=1}^{n}\sum_{k=1}^{K}\alpha_k\hat{\tau}_{ki}.
\end{align*}
Here, $$
W_{ki}=\frac{w_{\beta_k}(X_i)}{A_ie(X_i)+(1-A_i)(1-e(X_i))}
$$
and
$$
\hat{\tau}_{ki}=\left(\frac{A_i}{\frac{1}{n}\sum_{j=1}^{n}W_{kj}A_j}-\frac{1-A_i}{\frac{1}{n}\sum_{j=1}^{n}W_{kj}(1-A_j)}\right)W_{ki}Y_i.
$$
Then,
\begin{align}
\sqrt{n}(\hat{\tau}_P-\tau)&=\frac{1}{\sqrt{n}}\sum_{i=1}^{n}\sum_{k=1}^{K}\alpha_k\left[\left(\frac{A_i}{{\rm E}[W_{k}A]}-\frac{1-A_i}{{\rm E}[W_{k}(1-A)]}\right)W_{ki}Y_i-\tau\right. \nonumber \\
&\hspace{0.5cm}+\left(\frac{1}{\frac{1}{n}\sum_{j=1}^{n}W_{kj}A_j}-\frac{1}{{\rm E}[W_{k}A]}\right)A_iW_{ki}Y_i \nonumber \\
&\hspace{0.5cm}\left.-\left(\frac{1}{\frac{1}{n}\sum_{j=1}^{n}W_{kj}(1-A_j)}-\frac{1}{{\rm E}[W_{k}(1-A)]}\right)(1-A_i)W_{ki}Y_i\right]. \label{IF1}
\end{align}

For the second component of \eqref{IF1},
\begin{align*}
&\frac{1}{\sqrt{n}}\sum_{i=1}^{n}\left(\frac{1}{\frac{1}{n}\sum_{j=1}^{n}W_{kj}A_j}-\frac{1}{{\rm E}[W_{k}A]}\right)A_iW_{ki}Y_i\\
&\hspace{0.5cm}=-\left(\frac{\frac{1}{n}\sum_{i=1}^{n}A_iW_{ki}Y_i}{(\frac{1}{n}\sum_{j=1}^{n}W_{kj}A_j){\rm E}[W_{k}A]}\right)\sqrt{n}\left(\frac{1}{n}\sum_{i=1}^{n}W_{ki}A_i-{\rm E}[W_{k}A]\right)\\
&\hspace{0.5cm}=-\left(\frac{{\rm E}[AW_{k}Y]}{{\rm E}[W_{k}A]^2}\right)\frac{1}{\sqrt{n}}\sum_{i=1}^{n}\left(W_{ki}A_i-{\rm E}[W_{k}A]\right)+o_p(1)\\
&\hspace{0.5cm}=-\left(\frac{\theta^{0}_{1k}}{{\rm E}[w_{\beta_k}(X)]}\right)\frac{1}{\sqrt{n}}\sum_{i=1}^{n}\left(W_{ki}A_i-{\rm E}[w_{\beta_k}(X)]\right)+o_p(1).
\end{align*}
In the same way, for the third component of \eqref{IF1},
\begin{align*}
&\frac{1}{\sqrt{n}}\sum_{i=1}^{n}\left(\frac{1}{\frac{1}{n}\sum_{j=1}^{n}W_{kj}(1-A_j)}-\frac{1}{{\rm E}[W_{k}(1-A)]}\right)(1-A_i)W_{ki}Y_i\\
&\hspace{0.5cm}=-\left(\frac{\theta^{0}_{0k}}{{\rm E}[w_{\beta_k}(X)]}\right)\frac{1}{\sqrt{n}}\sum_{i=1}^{n}\left(W_{ki}(1-A_i)-{\rm E}[w_{\beta_k}(X)]\right)+o_p(1).
\end{align*}
Therefore, \eqref{IF1} becomes
\begin{align}
\sqrt{n}(\hat{\tau}_P-\tau)&=\frac{1}{\sqrt{n}}\sum_{i=1}^{n}\sum_{k=1}^{K}\alpha_k\left[\left(\frac{A_i}{{\rm E}[w_{\beta_k}(X)]}-\frac{1-A_i}{{\rm E}[w_{\beta_k}(X)]}\right)W_{ki}Y_i-\tau\right. \nonumber \\
&\hspace{0.5cm}-\left(\frac{\theta^{0}_{1k}}{{\rm E}[w_{\beta_k}(X)]}\right)\left(W_{ki}A_i-{\rm E}[w_{\beta_k}(X)]\right) \nonumber \\
&\hspace{0.5cm}\left.+\left(\frac{\theta^{0}_{0k}}{{\rm E}[w_{\beta_k}(X)]}\right)\left(W_{ki}(1-A_i)-{\rm E}[w_{\beta_k}(X)]\right)\right]+o_p(1). \label{IF2}
\end{align}

Focusing on $[\cdot]$ in \eqref{IF2},
\begin{align*}
&\left(\frac{A_i}{{\rm E}[w_{\beta_k}(X)]}-\frac{1-A_i}{{\rm E}[w_{\beta_k}(X)]}\right)W_{ki}Y_i-\tau -\left(\frac{\theta^{0}_{1k}}{{\rm E}[w_{\beta_k}(X)]}\right)\left(W_{ki}A_i-{\rm E}[w_{\beta_k}(X)]\right) \nonumber \\
&\hspace{1cm}+\left(\frac{\theta^{0}_{0k}}{{\rm E}[w_{\beta_k}(X)]}\right)\left(W_{ki}(1-A_i)-{\rm E}[w_{\beta_k}(X)]\right)\\
&\hspace{0.5cm}=\frac{W_{ki}A_i}{{\rm E}[w_{\beta_k}(X)]}(Y_i-\theta^{0}_{1k})-\frac{W_{ki}(1-A_i)}{{\rm E}[w_{\beta_k}(X)]}(Y_i-\theta^{0}_{0k})-(\tau-\tau_{\beta_k}).
\end{align*}
For $\sum_{k=1}^{K}\alpha_k\tau_{\beta_k}$, from \eqref{assu:poly} in the main manuscript,
$$
\sum_{k=1}^{K}\frac{1}{\bld{1}_K^{\top} \bar{P}_B\bld{1}_K}\left\{1-\sum_{k'=1}^{K}b_{k'k}\right\}\tau_{\beta_k}=\frac{\bld{1}_K^{\top}\bld{\tau}-\bld{1}_K^{\top}P_B\bld{\tau}}{\bld{1}_K^{\top} \bar{P}_B\bld{1}_K}=\frac{\bld{1}_K^{\top}\bar{P}_B\bld{\tau}}{\bld{1}_K^{\top} \bar{P}_B\bld{1}_K}=\tau.
$$
Whereas, for $\sum_{k=1}^{K}\alpha_k\tau$,
$$
\sum_{k=1}^{K}\frac{1}{\bld{1}_K^{\top} \bar{P}_B\bld{1}_K}\left\{1-\sum_{k'=1}^{K}b_{k'k}\right\}\tau=\tau\frac{\bld{1}_K^{\top}\bar{P}_B\bld{1}_K}{\bld{1}_K^{\top} \bar{P}_B\bld{1}_K}=\tau.
$$
Then,
$$
\sum_{k=1}^{K}\alpha_k(\tau_{\beta_k}-\tau)=0.
$$

From the discussion, \eqref{IF2} can be written as
\begin{align*}
\sqrt{n}(\hat{\tau}_P-\tau)&=\frac{1}{\sqrt{n}}\sum_{i=1}^{n}\sum_{k=1}^{K}\alpha_k\left[\frac{W_{ki}A_i}{{\rm E}[w_{\beta_k}(X)]}(Y_i-\theta^{0}_{1k})-\frac{W_{ki}(1-A_i)}{{\rm E}[w_{\beta_k}(X)]}(Y_i-\theta^{0}_{0k})\right]+o_p(1) \nonumber \\
&=:\frac{1}{\sqrt{n}}\sum_{i=1}^{n}\phi_i+o_p(1).
\end{align*}
Therefore,
$$
\sqrt{n}(\hat{\tau}_P-\tau)\stackrel{d}{\to}N(0,{\rm E}[\phi^2]).
$$

\subsection{Proof of Theorem \ref{coro1}}
Define
\begin{align*}
\widehat{W}_{ki}=W_{ki}(\hat{\zeta})&=\frac{\{e_i(\hat{\zeta})(1-e_i(\hat{\zeta}))\}^{\beta}}{A_ie_i(\hat{\zeta})+(1-A_i)(1-e_i(\hat{\zeta}))}\\
W_{ki}=W_{ki}(\zeta^{0})&=\frac{\{e_i(\zeta^{0})(1-e_i(\zeta^{0}))\}^{\beta}}{A_ie_i(\zeta^{0})+(1-A_i)(1-e_i(\zeta^{0}))}
\end{align*}
and
\begin{align}
\sqrt{n}(\hat{\tau}_P-\tau)&=\frac{1}{\sqrt{n}}\sum_{i=1}^{n}\sum_{k=1}^{K}\alpha_k\left[\left(\frac{A_i}{{\rm E}[W_{k}A]}-\frac{1-A_i}{{\rm E}[W_{k}(1-A)]}\right)\widehat{W}_{ki}Y_i-\tau\right. \nonumber \\
&\hspace{0.5cm}+\left(\frac{1}{\frac{1}{n}\sum_{j=1}^{n}\widehat{W}_{kj}A_j}-\frac{1}{{\rm E}[W_{k}A]}\right)A_i\widehat{W}_{ki}Y_i \nonumber \\
&\hspace{0.5cm}\left.-\left(\frac{1}{\frac{1}{n}\sum_{j=1}^{n}\widehat{W}_{kj}(1-A_j)}-\frac{1}{{\rm E}[W_{k}(1-A)]}\right)(1-A_i)\widehat{W}_{ki}Y_i\right]. \label{IF12}
\end{align}

Since $\hat{\zeta}$ is the MLE,
$$
\hat{\zeta}-\zeta^{0}=I(\zeta^{0})^{-1}\frac{1}{n}\sum_{i=1}^{n}S_i(\zeta^0)+o_p(1/\sqrt{n})
$$
under some regularity conditions.
Here,
$$
S_i(\zeta)=\frac{\partial}{\partial\zeta}\varphi_i(\zeta)(A_i-e_i(\zeta))
$$
is a score function, and $I(\zeta^{0})$ is the Fisher information.
Then,
\begin{align}
\widehat{W}_{ki}=W_{ki}+\dot{W}_{ki}^{\top}(\zeta^{0})(\hat{\zeta}-\zeta^{0})+o_p(1/\sqrt{n}), \label{WEE}
\end{align}
where $\dot{W}_{ki}(\zeta)=(\partial/\partial\zeta) W_{ki}(\zeta)$.
Specifically, using the fact that
$$
\frac{\partial}{\partial\zeta}e(\zeta)=e(\zeta)(1-e(\zeta))\frac{\partial}{\partial\zeta}\varphi(\zeta)=:e(\zeta)(1-e(\zeta))\dot{\varphi}(\zeta),
$$
$$
\dot{W}_{ki}(\zeta)=\frac{\partial}{\partial\zeta}\frac{\{e_i(\zeta)(1-e_i(\zeta))\}^{\beta_k}}{A_ie_i(\zeta)+(1-A_i)(1-e_i(\zeta))}=W_{ki}(\zeta)\left\{\beta_k(1-2e_i(\zeta))-(A_i-e_i(\zeta))\right\}\dot{\varphi}_i(\zeta).
$$

For $[\cdot]$ in \eqref{IF12},
\begin{align}
&\left(\frac{A_i}{{\rm E}[W_{k}A]}-\frac{1-A_i}{{\rm E}[W_{k}(1-A)]}\right)\widehat{W}_{ki}Y_i-\tau \nonumber+\left(\frac{1}{\frac{1}{n}\sum_{j=1}^{n}\widehat{W}_{kj}A_j}-\frac{1}{{\rm E}[W_{k}A]}\right)A_i\widehat{W}_{ki}Y_i \nonumber \\
&\hspace{0.5cm}-\left(\frac{1}{\frac{1}{n}\sum_{j=1}^{n}\widehat{W}_{kj}(1-A_j)}-\frac{1}{{\rm E}[W_{k}(1-A)]}\right)(1-A_i)\widehat{W}_{ki}Y_i \nonumber \\
&=\left(\frac{A_i}{{\rm E}[W_{k}A]}-\frac{1-A_i}{{\rm E}[W_{k}(1-A)]}\right)W_{ki}Y_i-\tau \nonumber \\
&\hspace{0.5cm}+\left(\frac{A_iY_i}{{\rm E}[W_{k}A]}-\frac{(1-A_i)Y_i}{{\rm E}[W_{k}(1-A)]}\right)(\widehat{W}_{ki}-W_{ki})\nonumber \\
&\hspace{0.5cm}-\left(\frac{A_iY_iW_{ki}}{(\frac{1}{n}\sum_{j=1}^{n}\widehat{W}_{kj}A_j){\rm E}[W_{k}A]}\right)\left(\frac{1}{n}\sum_{i=1}^{n}\widehat{W}_{ki}A_i-{\rm E}[W_{k}A]\right)\nonumber \\
&\hspace{0.5cm}-\left(\frac{A_iY_i(\widehat{W}_{ki}-W_{ki})}{(\frac{1}{n}\sum_{j=1}^{n}\widehat{W}_{kj}A_j){\rm E}[W_{k}A]}\right)\left(\frac{1}{n}\sum_{i=1}^{n}\widehat{W}_{ki}A_i-{\rm E}[W_{k}A]\right)\nonumber \\
&\hspace{0.5cm}+\left(\frac{(1-A_i)Y_iW_{ki}}{(\frac{1}{n}\sum_{j=1}^{n}\widehat{W}_{kj}(1-A_j)){\rm E}[W_{k}(1-A)]}\right)\left(\frac{1}{n}\sum_{i=1}^{n}\widehat{W}_{ki}(1-A_i)-{\rm E}[W_{k}(1-A)]\right)\nonumber \\
&\hspace{0.5cm}+\left(\frac{(1-A_i)Y_i(\widehat{W}_{ki}-W_{ki})}{(\frac{1}{n}\sum_{j=1}^{n}\widehat{W}_{kj}(1-A_j)){\rm E}[W_{k}(1-A)]}\right)\left(\frac{1}{n}\sum_{i=1}^{n}\widehat{W}_{ki}(1-A_i)-{\rm E}[W_{k}(1-A)]\right). \nonumber \\
\label{IF5}
\end{align}
From \eqref{WEE}, the second term of \eqref{IF5} is
\begin{align*}
&\frac{1}{\sqrt{n}}\sum_{i=1}^{n}\left[\left(\frac{A_iY_i}{{\rm E}[W_{k}A]}-\frac{(1-A_i)Y_i}{{\rm E}[W_{k}(1-A)]}\right)(\widehat{W}_{ki}-W_{ki})\right]\\
&\hspace{0.5cm}=\left(\frac{{\rm E}[(2A-1)Y\dot{W}_{k}^{\top}(\zeta^{0})]}{{\rm E}[w_{\beta_k}(X)]}\right)\sqrt{n}(\hat{\zeta}-\zeta^{0})+o_p(1).
\end{align*}
Similarly, from \eqref{WEE}, the third term of \eqref{IF5} is
\begin{align*}
&\frac{1}{\sqrt{n}}\sum_{i=1}^{n}\left[\left(\frac{A_iY_iW_{ki}}{(\frac{1}{n}\sum_{j=1}^{n}\widehat{W}_{kj}A_j){\rm E}[W_{k}A]}\right)\left(\frac{1}{n}\sum_{i=1}^{n}\widehat{W}_{ki}A_i-{\rm E}[W_{k}A]\right)\right]\\
&\hspace{0.5cm}=\left(\frac{{\rm E}[AYW_k]}{{\rm E}[w_{\beta_k}(X)]^2}\right)\left\{\frac{1}{\sqrt{n}}\sum_{i=1}^{n}\left(W_{ki}A_i-{\rm E}[W_{k}A]\right)+{\rm E}[A\dot{W}_{k}^{\top}(\zeta^{0})]\sqrt{n}(\hat{\zeta}-\zeta^{0})\right\}+o_p(1)\\
&\hspace{0.5cm}=\left(\frac{\theta^{0}_{1k}}{{\rm E}[w_{\beta_k}(X)]}\right)\frac{1}{\sqrt{n}}\sum_{i=1}^{n}\left(W_{ki}A_i-{\rm E}[w_{\beta_k}(X)]\right)\\
&\hspace{1cm}+\left(\frac{\theta^{0}_{1k}}{{\rm E}[w_{\beta_k}(X)]}\right){\rm E}[A\dot{W}_{k}^{\top}(\zeta^{0})]\sqrt{n}(\hat{\zeta}-\zeta^{0})+o_p(1),
\end{align*}
and the fifth term of \eqref{IF5} is
\begin{align*}
&\frac{1}{\sqrt{n}}\sum_{i=1}^{n}\left[\left(\frac{(1-A_i)Y_iW_{ki}}{(\frac{1}{n}\sum_{j=1}^{n}\widehat{W}_{kj}(1-A_j)){\rm E}[W_{k}(1-A)]}\right)\left(\frac{1}{n}\sum_{i=1}^{n}\widehat{W}_{ki}(1-A_i)-{\rm E}[W_{k}(1-A)]\right)\right]\\
&\hspace{0.5cm}=\left(\frac{\theta^{0}_{0k}}{{\rm E}[w_{\beta_k}(X)]}\right)\frac{1}{\sqrt{n}}\sum_{i=1}^{n}\left(W_{ki}(1-A_i)-{\rm E}[w_{\beta_k}(X)]\right)\\
&\hspace{1cm}+\left(\frac{\theta^{0}_{0k}}{{\rm E}[w_{\beta_k}(X)]}\right){\rm E}[(1-A)\dot{W}_{k}^{\top}(\zeta^{0})]\sqrt{n}(\hat{\zeta}-\zeta^{0})+o_p(1).
\end{align*}
The forth and sixth terms of \eqref{IF5} are $o_p(1)$ since there are so-called cross term.
Therefore, from \eqref{IF5},
\begin{align}
\sqrt{n}(\hat{\tau}_P-\tau)&=\frac{1}{\sqrt{n}}\sum_{i=1}^{n}\sum_{k=1}^{K}\alpha_k\left[\left(\frac{A_i}{{\rm E}[W_{k}A]}-\frac{1-A_i}{{\rm E}[W_{k}(1-A)]}\right)W_{ki}Y_i-\tau \right.\nonumber\\
&\hspace{0.5cm}+\left(\frac{{\rm E}[(2A-1)Y\dot{W}_{k}^{\top}(\zeta^{0})]}{{\rm E}[w_{\beta_k}(X)]}\right)I(\zeta^{0})^{-1}S_i(\zeta^0)\nonumber\\
&\hspace{0.5cm}-\left(\frac{\theta^{0}_{1k}}{{\rm E}[w_{\beta_k}(X)]}\right)\left(W_{ki}A_i-{\rm E}[w_{\beta_k}(X)]\right)\nonumber\\
&\hspace{1cm}-\left(\frac{\theta^{0}_{1k}}{{\rm E}[w_{\beta_k}(X)]}\right){\rm E}[A\dot{W}_{k}^{\top}(\zeta^{0})]I(\zeta^{0})^{-1}S_i(\zeta^0)\nonumber\\
&\hspace{0.5cm}+\left(\frac{\theta^{0}_{0k}}{{\rm E}[w_{\beta_k}(X)]}\right)\left(W_{ki}(1-A_i)-{\rm E}[w_{\beta_k}(X)]\right)\nonumber\\
&\hspace{1cm}+\left.\left(\frac{\theta^{0}_{0k}}{{\rm E}[w_{\beta_k}(X)]}\right){\rm E}[(1-A)\dot{W}_{k}^{\top}(\zeta^{0})]I(\zeta^{0})^{-1}S_i(\zeta^0)\right]+o_p(1)\nonumber\\
&=\frac{1}{\sqrt{n}}\sum_{i=1}^{n}\sum_{k=1}^{K}\alpha_k\left[\frac{W_{ki}A_i}{{\rm E}[w_{\beta_k}(X)]}(Y_i-\theta^{0}_{1k})-\frac{W_{ki}(1-A_i)}{{\rm E}[w_{\beta_k}(X)]}(Y_i-\theta^{0}_{0k})\right.\nonumber\\
&\hspace{0.5cm}+\frac{1}{{{\rm E}[w_{\beta_k}(X)]}}\left\{\left({\rm E}[AY\dot{W}_{k}^{\top}(\zeta^{0})]-\theta^{0}_{1k}{\rm E}[A\dot{W}_{k}^{\top}(\zeta^{0})]\right)\right.\nonumber\\
&\hspace{1cm}-\left.\left.\left({\rm E}[(1-A)Y\dot{W}_{k}^{\top}(\zeta^{0})]-\theta^{0}_{0k}{\rm E}[(1-A)\dot{W}_{k}^{\top}(\zeta^{0})]\right)\right\}I(\zeta^{0})^{-1}S_i(\zeta^0)\right]\nonumber\\
&\hspace{1.5cm}+o_p(1).
\label{IF6}
\end{align}
Since the first and second terms of \eqref{IF6} are the same as the influence function in Theorem \ref{theo1}, the theorem is proved.

\newpage
\section{Details of the approximation bias}\label{app:app_bias}
The condition \eqref{assu:poly2} in the main manuscript is sufficient to prove Theorem \ref{theo1} and \ref{coro1}.
Actually, for $\sum_{k=1}^{K}\alpha_k\tau_{\beta_k}$,
$$
\sum_{k=1}^{K}\frac{1}{\bld{1}_K^{\top} \bar{P}_B\bld{1}_K}\left\{1-\sum_{k'=1}^{K}b_{k'k}\right\}\tau_{\beta_k}=\frac{\bld{1}_K^{\top}\bld{\tau}-\bld{1}_K^{\top}P_B\bld{\tau}}{\bld{1}_K^{\top} \bar{P}_B\bld{1}_K}=\frac{\bld{1}_K^{\top}\bar{P}_B\bld{\tau}}{\bld{1}_K^{\top} \bar{P}_B\bld{1}_K}.
$$
Whereas, for $\sum_{k=1}^{K}\alpha_k\tau$,
$$
\sum_{k=1}^{K}\frac{1}{\bld{1}_K^{\top} \bar{P}_B\bld{1}_K}\left\{1-\sum_{k'=1}^{K}b_{k'k}\right\}\tau=\tau\frac{\bld{1}_K^{\top}\bar{P}_B\bld{1}_K}{\bld{1}_K^{\top} \bar{P}_B\bld{1}_K}.
$$
Then,
$$
\sum_{k=1}^{K}\alpha_k(\tau_{\beta_k}-\tau)=\frac{\bld{1}_K^{\top}\bar{P}_B(\bld{\tau}-\tau\bld{1}_K)}{\bld{1}_K^{\top} \bar{P}_B\bld{1}_K}=o\left(\frac{1}{\sqrt{n}}\right),
$$
and this term has no impact on the influence function for the PET method asymptotically.

Regarding the approximation bias \eqref{apperr1} in the main manuscript, we provide more details:
\begin{align*}
\gamma_{0}^{0}-\tau&=\frac{\bld{1}_K^{\top}\bar{P}_B}{\bld{1}_K^{\top}\bar{P}_B\bld{1}_K}\left(
\frac{{\rm E}\left[\left.h_{\beta}^{(q+1)}(X)\right|_{\beta=c_k(X)}\tau(X)\right]}{(q+1)!}\beta_k^{q+1}
\right)_{k=1,\dots,K}\\
&=\frac{1}{\bld{1}_K^{\top}\bar{P}_B\bld{1}_K}\sum_{k=1}^{K}\left\{1-\sum_{k'=1}^{K}b_{k'k}\right\}\frac{{\rm E}\left[\left.h_{\beta}^{(q+1)}(X)\right|_{\beta=c_k(X)}\tau(X)\right]}{(q+1)!}\beta_k^{q+1}.
\end{align*}
Here, we consider
\begin{align*}
&\left|\left\{1-\sum_{k'=1}^{K}b_{k'k}\right\}\frac{{\rm E}\left[\left.h_{\beta}^{(q+1)}(X)\right|_{\beta=c_k(X)}\tau(X)\right]}{(q+1)!}\beta_k^{q+1}\right|\\
&\hspace{0.5cm}=\left|1-\sum_{k'=1}^{K}b_{k'k}\right|\frac{\left|{\rm E}\left[\left.h_{\beta}^{(q+1)}(X)\right|_{\beta=c_k(X)}\tau(X)\right]\right|}{(q+1)!}\beta_k^{q+1}.
\end{align*}

Since $P_B$ is a projection matrix, $\sum_{k'=1}^{K}|b_{k'k}|<K$.
Then,
$$
\left|1-\sum_{k'=1}^{K}b_{k'k}\right|<1+K.
$$
Additionally, if
\begin{align}
\sup_{k=1,\dots,K}\left\{\frac{{\rm E}\left[\left.h_{\beta}^{(q+1)}(X)\right|_{\beta=c_k(X)}\right]}{(q+1)!}\right\}=:C^{K}_{q+1}<\infty, \label{assu:bound}
\end{align}
$$
\frac{\left|{\rm E}\left[\left.h_{\beta}^{(q+1)}(X)\right|_{\beta=c_k(X)}\tau(X)\right]\right|}{(q+1)!}<\sup_{x\in\mathcal{X}}\{|\tau(x)|\}C^{K}_{q+1}.
$$
Here, the assumption \eqref{assu:bound} provides a sufficient condition for bounding the Taylor remainder term.
Therefore, 
$$
|\gamma_{0}^{0}-\tau|<\frac{\sup_{x\in\mathcal{X}}\{|\tau(x)|\}K(1+K)}{\bld{1}_K^{\top}\bar{P}_B\bld{1}_K}C^{K}_{q+1}\beta_K^{q+1}.
$$
Note that $\beta_K<1$ from the definition of $D$. If $C^{K}_{q+1}< C_1$ and $\bld{1}_K^{\top}\bar{P}_B\bld{1}_K/K> C_2$ for some positive constants $C_1$ and $C_2$, then taking $\beta_{Kn}=O((nK)^{-1/q})$ is sufficient to ensure $\gamma_{0}^{0}-\tau=o(1/\sqrt{n})$.

Additionally, let us confirm that $\beta_{Kn}=O((nK)^{-1/q})$ implies $q_n \asymp O(\log(nK))$.
To see this, let
$$
\beta_K^{q_n+1}<\beta_K^{q_n}=:a_{nK}
$$
with fixed $\beta_K$.
Then, noting again that $\beta_K<1$,
$$
q_{nK}=\log_{\beta_K}(a_{nK})=\frac{-\log a_{nK}}{\log\frac{1}{\beta_K}}.
$$
Hence, choosing $q_n \asymp O(\log(nK))$ implies $\gamma_{0}^{0}-\tau=o(1/\sqrt{n})$.

\clearpage

\section{Additional simulation results}\label{app:add_sim}
\subsection{Full data generating mechanisms}
First, we generate six covariates $X_i=(X_{i1},\dots,X_{i6})^{\top}$.
To generate the covariates, we first generate $V_{ij}\sim N(0,1)$ ($j=1,\dots,6$) with $\text{Corr}(V_{ij},V_{ij'})=0.5$ ($j\neq j'$).
Based on the generated $V_{ij}$s, we define $X_{ij}\equiv V_{ij}$ ($j=1,2,3$) and $X_{ij}\equiv I\{V_{ij}>0\}$ ($j=4,5,6$).
The treatment assignment $A_i$ is generated from a Bernoulli distribution with
propensity score $e_i$:\ $A_i
\sim
\mathrm{Ber}(e_i)$ and $\mathrm{logit}(e_i)
=
\zeta_0+c X_i^\top\zeta$, where $\zeta=(0.15,0.3,0.3,-0.2,-0.25,-0.25)^\top$.
Here, $c=1$ corresponds to the strong overlap scenario, whereas $c=3$ corresponds to the limited overlap scenario. The intercept $\zeta_0$ is specified such that ${\rm E}[A_i]\approx 0.2$ in each scenario. The potential outcome is generated from the linear model
$$
Y_i(A_i)=
\Delta_i A_i+X_i^\top\xi+\varepsilon_i,
$$ 
where $\xi=(-0.5,-0.5,-1.5,0.8,0.8,1.0)^\top$ and $\varepsilon_i\sim N(0,1.5^2)$.
In the homogeneous setting, $\Delta_i\equiv0.75$.
In the heterogeneous setting, the conditional average treatment effect is given by $\Delta_i=0.25+0.5\left(X_{i1}+X_{i1}^2+X_{i2}\right)$. Under both settings, the true ATE is $\tau=0.75$.

\subsection{Details of the implementation of the PET method}
The implementation of the PET method follows the following steps. After calculating estimators for $\tau_{\beta}$, we conduct an ordinary least squares (OLS) procedure. Specifically, (1) for $\beta_k$ ($k=1,\dots,K$), obtain estimates of $\tau_{\beta}$; for instance, using the WATE estimators, denoted as $\hat{\tau}_{\beta}$; (2) specifying a polynomial regression model; for instance, when $q=2$,
$$
\hat{\tau}_{\beta_k}
=
\gamma_0+\gamma_{1}\beta_k+\gamma_{2}\beta^{2}_k+\epsilon_k.
$$ (3) Estimate $\boldsymbol{\gamma}$ by the OLS, and use $\hat{\gamma}_0$ ($=\hat{\tau}_P$) as the ATE estimator. In Step 2, we can use the \texttt{lm} function in R.

\subsection{Supplementary results for the main manuscript}
\begin{figure}[ht!]
\begin{center}
\begin{tabular}{c}
\includegraphics[width=16cm]{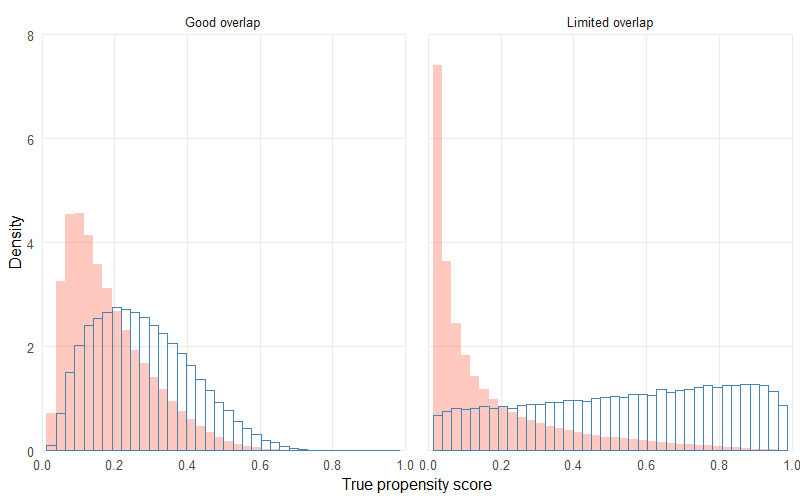}
\end{tabular}\caption{Histogram of the true propensity scores in 500,000 samples. Blue (outline):\ treatment group ($A=1$); Red (filled):\ control group ($A=0$).}
\label{zu3}
\end{center}
\end{figure}

Figure \ref{zu3} presents the distributions of the true propensity scores for each simulation scenario.

\vspace{0.5cm}

\begin{figure}[ht!]
\begin{center}
\begin{tabular}{c}
\includegraphics[width=17.5cm]{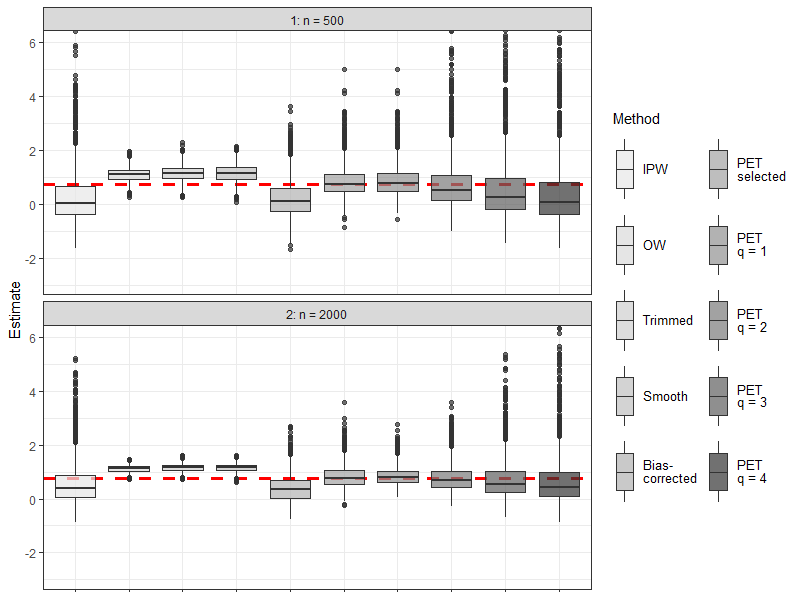}
\end{tabular}\caption{Boxplots of estimates from IPW-based estimators from the simulation experiment under the limited overlap scenario with the heterogeneous treatment effect. The number of iterations is 2,000, and the true ATE is $0.75$ (red dashed line). For Crump’s trimming method, \(\alpha\) is selected using the optimal cutoff implemented in the \texttt{PStrim} function in the \texttt{PSweight} package.
    For Yang’s smooth weighting method, \(\alpha\) is set equal to the cutoff selected for Crump’s method, with $\epsilon=10^{-2}$.
    For PET, $K=50$, $\beta_K=0.99$, and $\kappa=5/6$. Candidate values of $\beta_1$ are $\{0.44, 0.49, \dots, 0.69\}$ and $q$ are $\{1,2,3,4\}$.}
\label{fig1}
\end{center}
\end{figure}

\begin{figure}[ht!]
\begin{center}
\begin{tabular}{c}
\includegraphics[width=17.5cm]{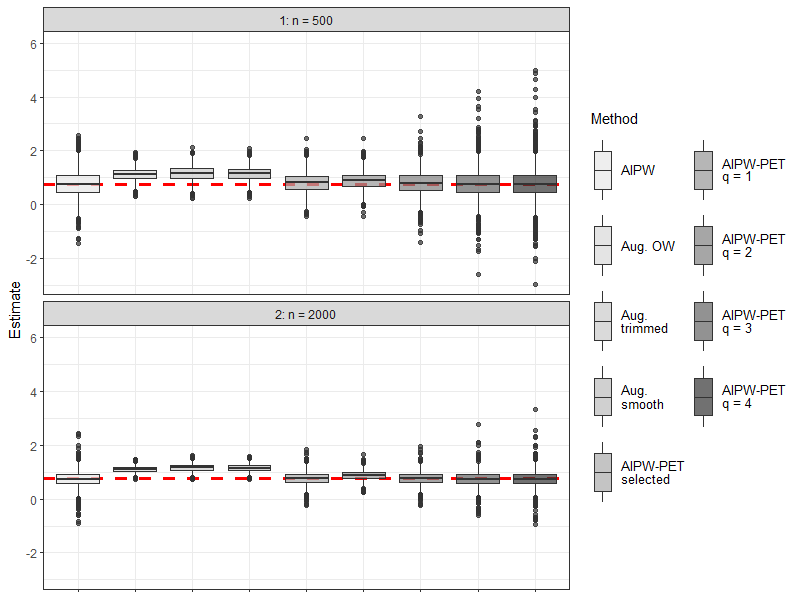}
\end{tabular}\caption{Boxplots of estimates from AIPW-based estimators from the simulation experiment under the limited overlap scenario with the heterogeneous treatment effect. The number of iterations is 2,000, and the true ATE is $0.75$ (red dashed line). For Crump’s trimming method, \(\alpha\) is selected using the optimal cutoff implemented in the \texttt{PStrim} function in the \texttt{PSweight} package.
    For Yang’s smooth weighting method, \(\alpha\) is set equal to the cutoff selected for Crump’s method, with $\epsilon=10^{-2}$.
    For AIPW-PET, $K=50$, $\beta_K=0.99$, and $\kappa=1$. Candidate values of $\beta_1$ are $\{0.44, 0.49, \dots, 0.69\}$ and $q$ are $\{1,2,3,4\}$.}
\label{fig2}
\end{center}
\end{figure}

\clearpage
Figures \ref{fig1} and \ref{fig2} present the boxplots of estimates from the IPW- and AIPW-based estimators in the simulation experiment under the limited overlap scenario with heterogeneous treatment effect setting, respectively.

\vspace{0.5cm}

\clearpage

\clearpage
\subsection{Good overlap scenario}
\begin{figure}[ht!]
\begin{center}
\begin{tabular}{c}
\includegraphics[width=17.5cm]{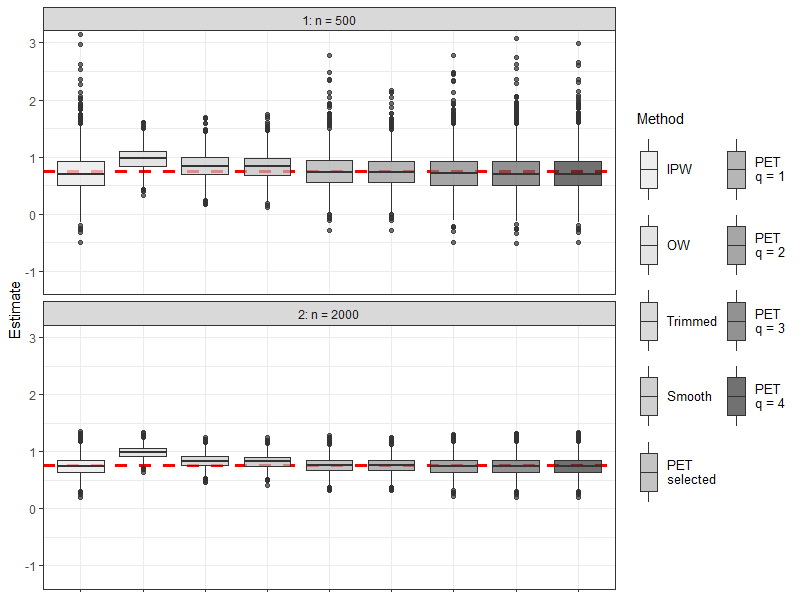}
\end{tabular}\caption{Boxplots of estimates from IPW-based estimators from the simulation experiment under the good overlap scenario with the heterogeneous treatment effect. The number of iterations is 2,000, and the true ATE is $0.75$ (red dashed line). For Crump’s trimming method, \(\alpha\) is selected using the optimal cutoff implemented in the \texttt{PStrim} function in the \texttt{PSweight} package.
    For Yang’s smooth weighting method, \(\alpha\) is set equal to the cutoff selected for Crump’s method, with $\epsilon=10^{-2}$.
    For PET, $K=50$, $\beta_K=0.99$, and $\kappa=5/6$. Candidate values of $\beta_1$ are $\{0.44, 0.49, \dots, 0.69\}$ and $q$ are $\{1,2,3,4\}$.}
\label{figA1}
\end{center}
\end{figure}

\begin{figure}[ht!]
\begin{center}
\begin{tabular}{c}
\includegraphics[width=17.5cm]{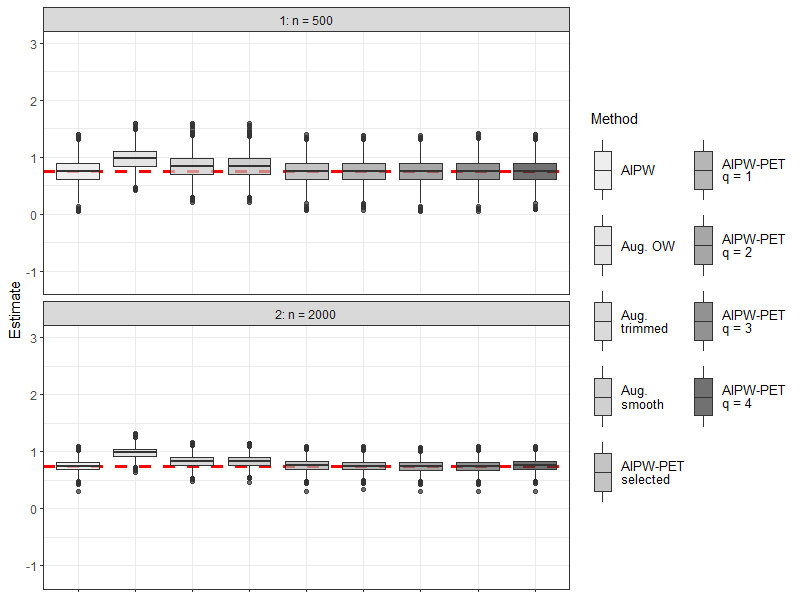}
\end{tabular}\caption{Boxplots of estimates from AIPW-based estimators from the simulation experiment under the good overlap scenario with the heterogeneous treatment effect. The number of iterations is 2,000, and the true ATE is $0.75$ (red dashed line). For Crump’s trimming method, \(\alpha\) is selected using the optimal cutoff implemented in the \texttt{PStrim} function in the \texttt{PSweight} package.
    For Yang’s smooth weighting method, \(\alpha\) is set equal to the cutoff selected for Crump’s method, with $\epsilon=10^{-2}$.
    For PET, $K=50$, $\beta_K=0.99$, and $\kappa=1$. Candidate values of $\beta_1$ are $\{0.44, 0.49, \dots, 0.69\}$ and $q$ are $\{1,2,3,4\}$.}
\label{figA2}
\end{center}
\end{figure}

The results are summarized in Table \ref{tab2}, and Figure \ref{figA1} and \ref{figA2}.
Here, since the number of subjects with propensity scores greater than $0.5$ in the control group is so limited, the bias-corrected estimator does not work in some cases; therefore, we do not summarize the results in this scenario.

The estimators for the ATE, including the standard IPW estimator, perform well compared with those under the limited overlap scenario.
The ATO, Crump's method, and Yang's method also show stable performance in terms of EmpSE with small biases.
For the PET methods, both the bias and EmpSE are small, indicating reasonable performance even under the good overlap scenario.
Importantly, the hyperparameter selection does not substantially affect the PET results, at least compared with the limited overlap scenario.

The AIPW-PET results are more similar to those of the standard AIPW estimator than under the limited overlap scenario.
This similarity may provide some rationale for the derivation of the augmented influence function in Section \ref{sec:aug} in the main manuscript; the (asymptotic) behavior of the augmented influence function may become similar to the standard efficient influence function for the ATE $\tau$.
From these results, the PET methods can potentially serve as alternatives to the standard estimators for the ATE.

\newpage
\begin{table}[ht!]
\begin{center}
\caption{Results of the simulation experiment under the good overlap scenario with the heterogeneous treatment effect. The number of iterations is 2,000.}\label{tab2}
    \scalebox{0.9}{$
    \begin{tabular}{cc|ccccc|ccccc}
    \midrule
       \multicolumn{10}{l}{\textbf{IPW estimators}} \\
       \addlinespace[0.3em]
       \hline
       Method & $q$ & \multicolumn{5}{c|}{$n=500$} & \multicolumn{5}{c}{$n=2000$} \\ 
        & & $|$bias$|$ & EmpSE & RMSE & $\frac{\mathrm{EstSE}}{\mathrm{EmpSE}}$ & CP 
        & $|$bias$|$ & EmpSE & RMSE & $\frac{\mathrm{EstSE}}{\mathrm{EmpSE}}$ & CP \\ \hline
          IPW & -- & 0.022 & 0.361 & 0.361 & 0.849 & 91.2 & 0.005 & 0.166 & 0.166 & 0.962 & 93.2 \\
          OW & -- & 0.219 & 0.195 & 0.293 & 0.990 & 79.6 & 0.231 & 0.099 & 0.251 & 0.976 & 34.4 \\
      Trimmed & -- & 0.094 & 0.229 & 0.247 & 0.953 & 92.1 & 0.080 & 0.117 & 0.141 & 0.947 & 87.5 \\
       Smooth & -- & 0.081 & 0.230 & 0.244 & 1.219 & 97.6 & 0.074 & 0.118 & 0.140 & 1.037 & 91.1 \\
          PET & selected & 0.004 & 0.306 & 0.306 & 0.899 & 93.5 & 0.005 & 0.144 & 0.144 & 0.984 & 94.6 \\
              & 1 & 0.000 & 0.293 & 0.293 & 0.924 & 93.6 & 0.003 & 0.141 & 0.141 & 0.985 & 94.6 \\
              & 2 & 0.021 & 0.343 & 0.344 & 0.874 & 91.4 & 0.008 & 0.160 & 0.160 & 0.968 & 93.4 \\
              & 3 & 0.021 & 0.361 & 0.362 & 0.846 & 91.0 & 0.005 & 0.165 & 0.165 & 0.962 & 93.2 \\
              & 4 & 0.021 & 0.364 & 0.365 & 0.797 & 90.1 & 0.005 & 0.166 & 0.166 & 0.916 & 92.1 \\
       \hline
       \midrule
       \multicolumn{10}{l}{\textbf{AIPW estimators}} \\
       \addlinespace[0.3em]
       \hline
       Method & $q$ & \multicolumn{5}{c|}{$n=500$} & \multicolumn{5}{c}{$n=2000$} \\ 
        & & $|$bias$|$ & EmpSE & RMSE & $\frac{\mathrm{EstSE}}{\mathrm{EmpSE}}$ & CP 
        & $|$bias$|$ & EmpSE & RMSE & $\frac{\mathrm{EstSE}}{\mathrm{EmpSE}}$ & CP \\ \hline
          AIPW & -- & 0.005 & 0.208 & 0.208 & 0.957 & 94.2 & 0.001 & 0.106 & 0.106 & 0.949 & 93.2 \\
          Aug.\ OW & -- & 0.222 & 0.187 & 0.290 & 0.977 & 77.0 & 0.231 & 0.098 & 0.251 & 0.943 & 29.7 \\
      Aug.\ trimmed & -- & 0.087 & 0.202 & 0.221 & 0.964 & 91.9 & 0.075 & 0.105 & 0.129 & 0.936 & 86.8 \\
       Aug.\ smooth & -- & 0.088 & 0.198 & 0.217 & 0.967 & 91.9 & 0.075 & 0.102 & 0.127 & 0.943 & 86.5 \\
          AIPW-PET & selected & 0.000 & 0.204 & 0.204 & 0.966 & 94.7 & 0.006 & 0.106 & 0.106 & 0.947 & 93.1 \\
              & 1 & 0.000 & 0.200 & 0.200 & 0.976 & 94.8 & 0.002 & 0.103 & 0.103 & 0.953 & 93.2 \\
              & 2 & 0.009 & 0.207 & 0.207 & 0.967 & 94.8 & 0.005 & 0.106 & 0.106 & 0.951 & 93.2 \\
              & 3 & 0.007 & 0.208 & 0.209 & 0.963 & 94.6 & 0.002 & 0.106 & 0.106 & 0.950 & 93.0 \\
              & 4 & 0.003 & 0.207 & 0.207 & 0.961 & 94.5 & 0.007 & 0.106 & 0.106 & 0.946 & 93.1 \\
       \hline
    \end{tabular}
       $}
\end{center}
    {\footnotesize{
    Notes:\ EmpSE, empirical standard error; RMSE, root mean squared error; EstSE, mean estimated standard error; CP, coverage probability of the 95\% confidence interval.
    For Crump’s trimming method, \(\alpha\) is selected using the optimal cutoff implemented in the \texttt{PStrim} function in the \texttt{PSweight} package.
    For Yang’s smooth weighting method, \(\alpha\) is set equal to the cutoff selected for Crump’s method, with $\epsilon=10^{-2}$.
    For PET, $K=50$, $\beta_K=0.99$, and $\kappa=5/6$ (for the PET) or $\kappa=1$ (for the AIPW-PET). Candidate values of $\beta_1$ are $\{0.44, 0.49, \dots, 0.69\}$ and $q$ are $\{1,2,3,4\}$.
    For PET and AIPW-PET, confidence intervals use the variance estimators in \eqref{avar2} and \eqref{avar3}, respectively. For the remaining IPW estimators, robust variance estimators accounting for propensity score estimation are used.
    }}
\end{table}

\clearpage

\clearpage

\clearpage
\subsection{Homogeneous treatment effect setting}
Returning to the original \cite{li2019addressing} setting, we consider the homogeneous treatment effect setting:\ $\Delta_i\equiv0.75$.
Here, we only consider the limited overlap scenario ($c=3$).

The results are summarized in Table \ref{tab:app5}.
Except for the bias-corrected and PET methods, the overall tendencies are consistent with the results of \cite{li2019addressing}.
The bias-corrected method shows the same tendency as in the heterogeneous treatment effect setting in the main manuscript.
Under the homogeneous treatment effect setting, since the WATEs are the same for all values of $\beta$, $\beta=1$, or the value nearest to $\beta=1$, is the best choice for the PET method \citep{crump2009dealing,li2018balancing}.
However, even when $q=1$, the performance of the ATO, trimmed, or smoothed estimators is better than that of PET.
One reason is that PET involves an unnecessarily complex estimation procedure compared with these methods under this setting.
Therefore, when a homogeneous treatment effect is considered, PET is not an initial choice for estimating the ATE; ATO is the best choice in terms of efficiency.

\newpage
\begin{table}[ht!]
\begin{center}
\caption{Results of the simulation experiment with the homogeneous treatment effect. The number of iterations is 2,000.}\label{tab:app5}
    \scalebox{0.9}{$
    \begin{tabular}{cc|ccccc|ccccc}
    \midrule
       \multicolumn{10}{l}{\textbf{IPW estimators}} \\
       \addlinespace[0.3em]
       \hline
Method & $q$ & \multicolumn{5}{c|}{$n=500$} & \multicolumn{5}{c}{$n=2000$} \\ 
        & & $|$bias$|$ & EmpSE & RMSE & $\frac{\mathrm{EstSE}}{\mathrm{EmpSE}}$ & CP 
        & $|$bias$|$ & EmpSE & RMSE & $\frac{\mathrm{EstSE}}{\mathrm{EmpSE}}$ & CP \\ \hline
IPW
& -- 
& 0.506 & 1.180 & 1.284 & 0.470 & 50.8
& 0.202 & 0.850 & 0.873 & 0.599 & 67.2 \\

OW
& --
& 0.003 & 0.225 & 0.225 & 0.964 & 93.7
& 0.003 & 0.107 & 0.107 & 1.019 & 95.2 \\

Trimmed
& --
& 0.005 & 0.274 & 0.274 & 0.922 & 93.0
& 0.003 & 0.129 & 0.130 & 0.998 & 94.7 \\

Smooth
& --
& 0.010 & 0.314 & 0.314 & 1.170 & 97.0
& 0.003 & 0.152 & 0.152 & 1.062 & 96.2 \\

Bias-corrected
& --
& 0.542 & 0.762 & 0.935 & 0.886 & 91.1
& 0.327 & 0.566 & 0.653 & 1.040 & 85.2 \\

PET
& selected
& 0.063 & 0.630 & 0.633 & 0.776 & 84.8
& 0.003 & 0.461 & 0.461 & 0.911 & 86.8 \\

& 1
& 0.034 & 0.610 & 0.611 & 0.797 & 86.7
& 0.008 & 0.380 & 0.380 & 0.855 & 87.2 \\

& 2
& 0.050 & 1.134 & 1.135 & 0.596 & 73.8
& 0.008 & 0.573 & 0.573 & 0.841 & 84.5 \\

& 3
& 0.140 & 1.632 & 1.638 & 0.471 & 63.3
& 0.032 & 0.842 & 0.842 & 0.724 & 78.6 \\

& 4
& 0.286 & 1.759 & 1.782 & 0.389 & 53.9
& 0.070 & 1.044 & 1.046 & 0.588 & 71.0 \\
       \hline
    \end{tabular}
       $}
\end{center}
    {\footnotesize{
    Notes:\ EmpSE, empirical standard error; RMSE, root mean squared error; EstSE, mean estimated standard error; CP, coverage probability of the 95\% confidence interval.
    For Crump’s trimming method, \(\alpha\) is selected using the optimal cutoff implemented in the \texttt{PStrim} function in the \texttt{PSweight} package.
    For Yang’s smooth weighting method, \(\alpha\) is set equal to the cutoff selected for Crump’s method, with $\epsilon=10^{-2}$.
    For PET, $K=50$, $\beta_K=0.99$, and $\kappa=5/6$. Candidate values of $\beta_1$ are $\{0.44, 0.49, \dots, 0.69\}$ and $q$ are $\{1,2,3,4\}$.
        For the bias-corrected method, a confidence interval based on studentized subsampling is used.
    For PET, confidence intervals use the variance estimators in \eqref{avar2}. For the remaining IPW estimators, robust variance estimators accounting for propensity score estimation are used.
    }}
\end{table}

\clearpage
\section{Additional results for the real data application}\label{app:realdata}
\subsection{Impact of right heart catheterization on mortality}
\begin{figure}[ht!]
\begin{center}
\begin{tabular}{c}
\includegraphics[width=15cm]{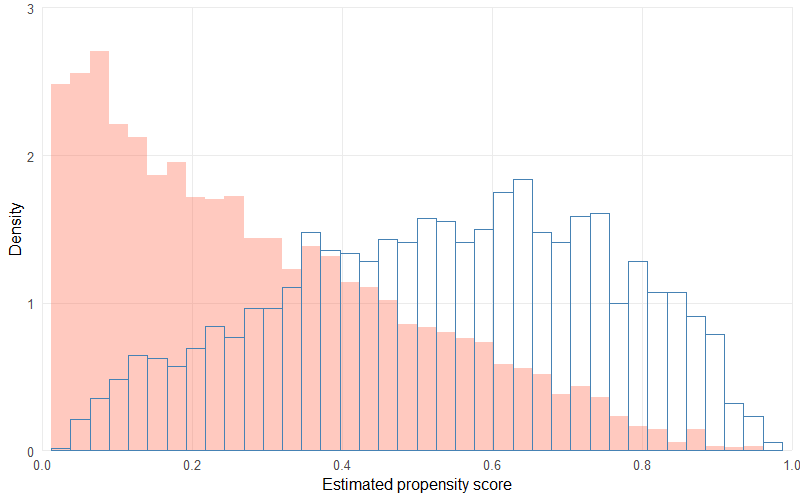}
\end{tabular}\caption{Histogram of the estimated propensity scores in the RHC data. Blue (outline):\ treatment group ($A=1$); Red (filled):\ control group ($A=0$).}
\label{zu_ps}
\end{center}
\end{figure}

Figure \ref{zu_ps} presents the distributions of the estimated propensity scores in the RHC data.

\vspace{0.5cm}

\newpage
\subsection{Impact of smoking habits on forced expiratory volume}
\begin{figure}[ht!]
\begin{center}
\begin{tabular}{c}
\includegraphics[width=15cm]{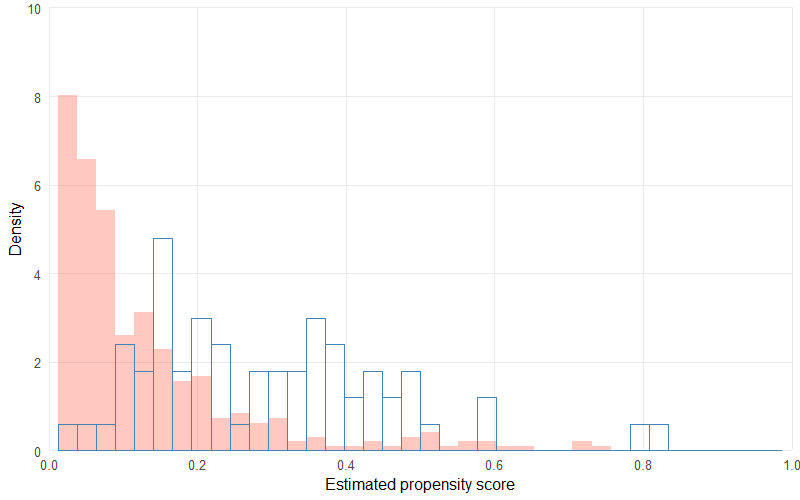}
\end{tabular}\caption{Histogram of the estimated propensity scores in the FEV data. Blue (outline):\ treatment group ($A=1$); Red (filled):\ control group ($A=0$).}
\label{zu_ps_fev}
\end{center}
\end{figure}

Figure \ref{zu_ps_fev} presents the distributions of the estimated propensity scores in the FEV data.
\vspace{0.5cm}

\end{document}